# ZnO-based Semiconductors and Structures for Transistors, Optoelectronic Devices and Sustainable Electronics

Darragh Buckley[1], Alex Lonergan[1], and Colm O'Dwyer[1,2,3]

[1]*School of Chemistry, University College Cork, Cork, T12 YN60, Ireland*
[2]*AMBER@CRANN, Trinity College Dublin, Dublin 2, Ireland*
[3]*Environmental Research Institute, University College Cork, Lee Road, Cork T23 XE10, Ireland*

**Abstract**

Metal oxide thin films are of great interest in scientific advancement, particularly semiconductor thin films in transistors and in a wide range of optoelectronic applications. Many metal oxide thin films attract interest for their electronic bandgap, charge carrier mobility, optical opacity, luminescence, low cost, relative abundance and environmentally-friendly production. Additionally, these properties are often tuneable via particle size, film density, surface morphology, film deposition, growth method, hetero-interface engineering or ion-doping. Zinc oxide as a n-type semiconducting metal oxide is material of great interest owing to its intrinsically wide direct bandgap, high electron mobility, relatively high exciton binding energy, high optical transparency, demonstrated metal-ion doping optoelectronic effects, a range of different particle morphologies and deposition methods, photoluminescence ability, low cost and a variety of existing green synthesis methods. Here, these aspects of zinc oxide and some related oxides are reviewed, focusing on how the unique properties of this metal oxide make it suitable for a range of different applications from thin film transistors, high mobility oxide interfaces, transparent conductive oxides, photoanodes photodetectors, chemical sensors, photocatalysts, superlattice electronics and more. The properties and deposition methods and their impact on functionality will be discussed alongside their role in sustainable optoelectronics for future devices.

**Introduction**

Metal oxides form a backbone material set for many applications from batteries and energy storage or conversion, to magnetics, catalysis and a myriad of important uses in electronics and optoelectronics. As will be described in this review, the ability to controlled stoichiometry, structure and thin film deposition or coating are key capabilities for their application and in the discovery of new properties. More recently, the use of more complex oxides, heterostructures of oxides, intricate nanoscale structures, and the ability to modify a very large range of even simple transition metal oxides in semiconductor form has driven a lot of exciting physics, electronic engineering and material science. Metal oxides in semiconducting form provide a lot of flexibility in preparation and physical or chemical qualities in comparison to some single crystal semiconducting materials such as III-N, III-V or Group 14 compounds (Si, Ge etc.). Metal oxide materials provide an opportunity for important research into the variety of useful properties and phenomena that are possible through tailored deposition for specific applications. Oxides provide an open-ended challenge to identify the best candidates for developing and emerging technologies, based on the current breadth of research completed to date.

The potential applications for metal oxides also benefits when materials are deposited, and investigated, at the nano or micro scale. A variation in the useful properties and characteristics from bulk materials to few-micrometer thick films, to homo- or heterointerfaces or to single atom layers greatly broadens the possibilities for a group of materials and their potential application. The interest in oxide materials covers a broad range of research areas including nanoelectronics, photovoltaics, solar cells, sensors and environmental uses. Coupled with the natural abundance of oxides of relatively abundant transition metals among others, the characteristics that make metal oxides an attractive area for researching technological growth include high carrier mobility in electronics, transparency to visible light, low processing temperatures, controllable conductivity and a variety of suitable deposition methods.

This review is primarily concerned with the use of metal oxides in thin film form, with a particular focus on recent advances in layers of films of zinc oxide (ZnO) based metal oxides for optoelectronic and related applications. These films, grown via a multitude of techniques, can be comprised of various structures and morphologies, lending to different electrical, optical and physical properties that are advantageous in particular areas of research. This inherent diversity in the uses of metal oxide thin films is one of the main driving factors for the breadth of the research conducted in the area in recent years.

Oxide semiconductors have garnered significant research interest over the last two decades, primarily due to work carried out by Hosono et al. showing the use of transparent conducting oxides (TCOs) for use in thin-film transistor (TFT) technologies.(1) Oxide TFTs present a number of advantages over the established hydrogenated amorphous silicon (a-Si:H)(2) thin film transistors in that they are capable of providing equal or greater field effect mobility while retaining useful properties such as high transparency to light and low-temperature processing that does not rely on vacuum-based methods. However, oxide electronic materials are becoming increasingly important in a wide range of applications including transparent electronics(3), optoelectronics(4), magnetoelectronics(5), photonics(6), spintronics(7), thermoelectrics(8), piezoelectrics(9), power harvesting(10), hydrogen storage(11) and environmental waste management.(12)

Transparent conductive oxides are an important subsection of oxide materials due to their ability to have controllable conductivity and carrier mobility, while maintaining high optical transparency.(13) This trade-off between optical and electrical conductivity around the plasma frequency has motivated research into controlled and graded porosity in materials and structures to offset transparency loss in conductive oxides.(14-18) Metal oxide TCOs leads to the use of oxide semiconductors in many key optoelectronic devices such as TFTs(19), photovoltaics(20), solar cells(21) and electrochromics.(22)

As an important topic of research, metal oxide electronics are the subject of much review,(2, 4, 23-31) where the challenges being addressed in the synthesis of cutting edge materials along with advances in the characterization techniques are highlighted in an effort to further improve the development of oxide technologies for modern applications. As a semiconducting metal oxide, ZnO has been the subject of much scientific research with many reports on specific implementations or covering a range of different applications for this versatile material(32) (33) (34) (35). This review aims to highlight the key results and the most recent advances in the area of metal oxide thin film by solution deposition, and the myriad applications that have resulted from newer methods of substrate engineering, ZnO and related materials processing(36), optoelectronics and the advent of greener synthesis and sustainable electronics. Due to the ever growing and vast scope of the ongoing research in this area, the review is presented with a focus on advancements using the preparation and modification of zinc oxide-based materials and devices. Additionally, advancements in green chemistry synthesis of ZnO-based nanomaterials are discussed for the future of sustainable optoelectronic devices fabricated from natural materials to meet future technological demand while minimizing the impact on the environment.

**Thin Film Applications**

Thin film technologies look to explore the benefit of dramatically increasing the surface to volume ratio of materials. This allows for the exploitation of interesting electrical and optical properties of nanoscale metal oxides that synergize with the ever growing need for more compact electronic devices.

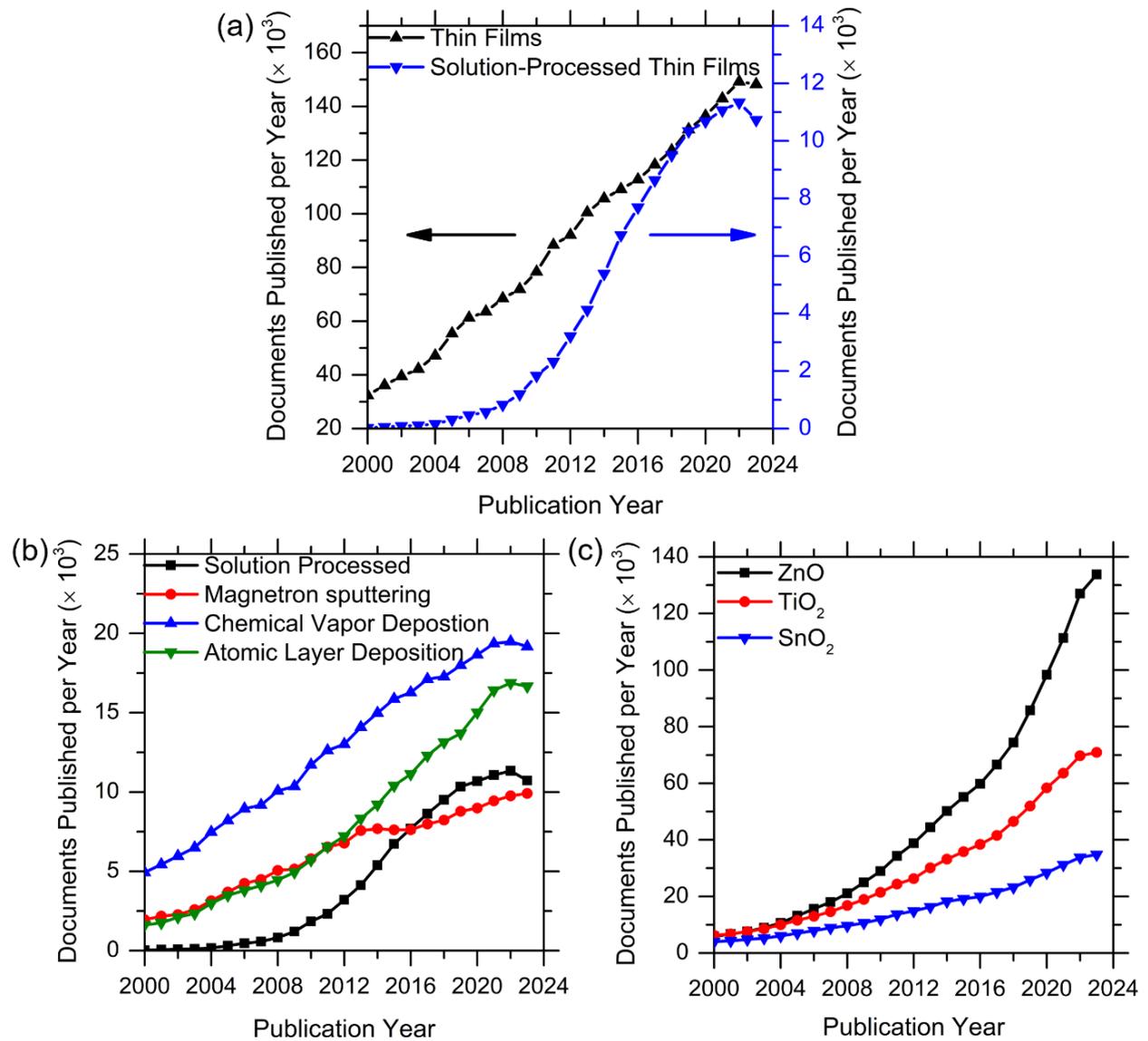

**Figure 1.** (a) Number of documents published per year relating to thin films and solution processed thin films from publication years 2000-2023. (b) Comparison of thin film preparation methods using the number of documents published per year with search terms containing the preparation methods and the term thin film. (c) Comparison of the number of documents published per year for common metal oxides reported in the literature. All data was obtained by via a SCOPUS search for the specific terms in the plot legends.

Indium-free alternatives for conductive oxide films are becoming increasingly sought after due to the predicted future scarcity, availability and cost of indium(37) (38). Indium tin oxide (ITO) is commonly used as the industry standard(38) for transparent conductive oxide films yet it does face issues with the brittleness of the film layer, particularly for applications in flexible electronic devices(39) (40). Some researchers are currently investigating routes to recycle indium from flat panel display devices(41) (42) (43), while others are actively searching for transparent conductive oxide alternatives using more abundant raw materials or with increased compatibility for flexible electronics(44) (45) (46). Zinc oxide and its doped counterparts are extensively investigated as indium-free alternatives for metal oxide electronics including solar cells(47) (48), light emitting diodes (LEDs)(49) (50) and also display technologies(51) (52). For the latter application and for the area of macroelectronics, thin film transistor (TFT) devices are the building-block structures with progress into solution processed thin film preparation methods and investigating the optical/electrical properties of materials accelerating research interest in metal oxides for display technologies(53) (54).

Due the diverse range of materials that can be deposited as thin films, and the ever-growing number of applications, the area of thin film technologies is continuing to grow in popularity over time. Figure 1 (a) highlights that within the last two decades there has been significant growth in the research conducted on thin films and solution processed thin films, which highlights the focus on the need to move to large-scale, cost-effective and vacuum-free manufacturing processes for a range of thin film applications(55) (56). Figure 1 (b) shows a comparison between research articles pertaining to thin film preparation techniques, showing a general trend of growth for thin films prepared via magnetron sputtering, chemical vapor deposition, atomic later deposition and solution processing. In particular, solution processing preparation has seen a marked increase with an onset in the mid-2000s.

Figure 1 (c) compares the number of articles published for common metal oxides in the literature. In comparison to $TiO_2$ and $SnO_2$ research, ZnO research has seen a marked increase in research initiatives with a significant increase beginning in the mid-2000s, coinciding with the growing popularity of solution processing of thin films around this time period. As a result of the large volume of work in the area of zinc oxide thin films, recent examples of the application of its optical and electrical properties in a number of different areas can be found. Patil et al. present work on a $NO_2$ gas sensor device based on ZnO thin films formed via spin coating deposition of a zinc acetate in ethanol and m-cresol precursor solution.(57) Films were annealed at 400 °C and produced highly crystalline films which demonstrated fast response and recovery time to the detection of $NO_2$ gas. More recently, ZnO nanocrystals were spin coated in conjunction with $CsPbBr_3$ quantum dots by Liu et al. to achieve a blue perovskite LED with an emission of 470 nm light at a record level of external quantum efficiency of blue light in perovskite LEDs(58).

Metal oxide thin films can be used as different components within the same devices, depending on the chosen material and deposition methods. In the area of electronic devices, Huang et al. have recently demonstrated a zinc oxide-based resistive random access memory (RRAM) device, with I-V behaviour that is shown in Figure 2 (a).(59) The amorphous ZnO films were deposited via RF sputtering and exhibited high non-volatile resistive switching performance. The use of metal oxide films as multiple components in devices is exemplified in solar cell research, with ZnO metal oxides in particular, where modified ZnO materials have been utilized as the electrode contacts in these thin film solar cell configurations. Sharma et al. report on the use of low-roughness, layered ZnO-Ag-ZnO transparent conducting electrodes deposited via a RF sputtering method (Figure 2 (b)), which demonstrate low optical and electrical response under irradiation, for use in solar cell devices and other radiation-harsh optoelectronic applications.(60) Research related to thin films is reportedly dominated by thin film solar cells.(61) Eisner et al. have recently shown inorganic colloidal quantum dot solar cells with the incorporation of a solution process In$_2$O$_3$/ZnO electron transport layer, demonstrating very high power conversion efficiencies, show in Figure 2 (c).(62)

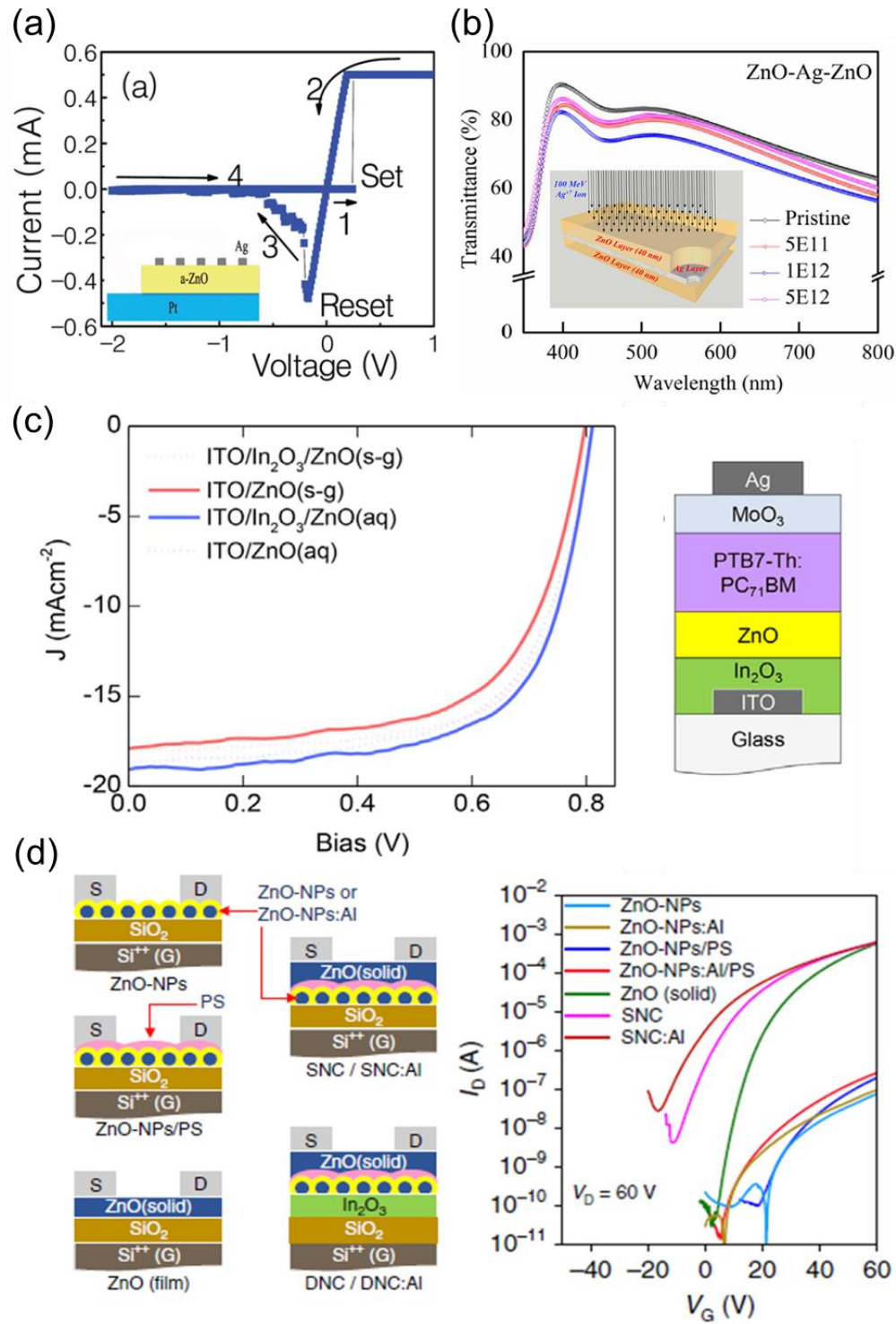

**Figure 1.** Examples of different thin film applications of zinc oxide-based materials (a) Random access memory device showing characteristic bi-polar *I-V* curve.(59) (b) ZnO-Ag-ZnO transparent conducting electrode schematic, inset, on transmittance spectra in the visible region.(60) (c) *J-V* characteristics (left) of solar cells of using metal oxide thin film architecture

(right), measured under simulated solar illumination.(62) (d) Schematic of the device structures for transistors based on ZnO-NPs and ZnO-NPs/PS (left) with corresponding transfer characteristics (right).(63)

An area where thin film research is also highly focused is the area of thin film transistors. This work is driven by the seminal work from Hosono et al.(1) and continues to be a popular field of research due to the broad scope of materials and deposition methods that can be incorporated into the process of TFT device fabrication. Work by the Anthopoulos group shows multilayer channel TFT architectures incorporating polystyrene layers with ZnO nanoparticles deposited on pseudo-2D metal oxide layers, shown in Figure 2 (d).(63) These devices build off the significant body of work from this research group in the area of metal oxide TFT devices from solution methods(64-70) to show high electron mobility and operational stability. More recently, they have developed biosensors for metabolite (uric acid and vitamin D3) detection in human saliva with a solution processed (spin coated) heterojunction transistor of $In_2O_3$ and ZnO functionalized with either an enzyme or antibody at the core of its operation, facilitating health diagnostic capabilities(71).

These examples highlight some of the areas where metal oxide thin film technologies are employed for their favorable electronic and/or optical properties. Other thin film applications include dielectrics(72), thermochromics and electrochromics(73), antireflective coatings(74, 75), photocatalysis(76) (77), energy storage(78) (79) UV photodetectors(80), refractive index sensors(81) (82) and a range of chemical(83) (84) and biosensors(85) (86) (87). The continued development of metal oxide thin film materials, through the study and understanding of the characteristics of various oxide materials deposited through numerous solution-based methods, is crucial for the improvement of many modern technologies.

**Wide Band Gap Materials**

Transparent conductive oxides (TCOs) are a core material class in optoelectronics due to their ability to have controllable conductivity and carrier mobility, while maintaining high optical transparency.(13) Zinc oxide and its doped counterparts such as Al:ZnO (AZO) are researched for use in many key optoelectronic devices such as TFTs(19), photovoltaics(20), solar cells(21) and electrochromics.(22) In the search for high-mobility, cheap, optoelectronic metal oxides, ZnO has found much popularity due to its many desirable properties.

**Zinc Oxide**

Zinc oxide attracted the initial attention in the area of optoelectronics due to its wide, direct band gap ($E_g \sim 3.3$ eV at 300 K),(26) a conductivity influenced by defects and cation/anion vacancies, its ability to alloy with other metal in oxide form and a crystal lattice that facilitates interstitial doping.(88)

ZnO has been used in the semiconductor industry for a large range of devices and applications, and has been extensively studied. There is a large volume of research into the varying properties of ZnO with respect to the growth methods and the resulting nanostructures that are possible(89) (90). ZnO is unique among the metal oxide/TCO family of materials due to the vast configurations which can be formed through simple and low-cost processes(91) (92) (93). The semiconducting and wide-band gap properties of ZnO make it an attractive material for use in a number of fields. However, coupling those properties with the ability to control the growth of specific nanostructures gives ZnO a range of characteristics that have commercial uses in fields such as energy harvesting(94) (95), catalysis(96) (97), electronics/optoelectronics(98) (99), sensors(96) (100) and display technologies(101) (102).

The driving force behind ZnO research in TFTs is to source a metal oxide material that has high-mobility and can be deposited on several amorphous and flexible substrates. Transparency in the visible is advantageous for some applications, particularly for "invisible electronic circuits" in future of display technology and electronics, with the proviso that lower-temperature processing and control of conduction/mobility is improved.(103) However, good control over solution processing is paramount for high quality electronic material grade ZnO and its alloyed or doped counterparts, particularly when grown at low temperature(104). At just above the decomposition temperatures for the organic moieties of Zn-based precursors, effective removal of these species is critical for crystallization and defined stoichiometry. It is also imperative that high crystal quality epitaxial-like thin films can be formed from solution processing to compete with physical deposition methods. Oxygen vacancy formation within ZnO can sometimes be balanced at higher temperatures in $O_2$ atmospheres, where oxidation annihilates oxygen vacancy density, but it has yet to be determined if surface ionic vacancy formation is modified during lower temperature crystallization to a high electronic quality thin film.(105) Other interstitial and vacancy defects are also formed in ZnO and understanding the nature of these during crystallization of thin films at lower temperatures remains a challenge since stoichiometry and composition determine conductivity and free carrier mobility.(106)

ZnO can be grown as a polycrystalline material at low temperatures or even room temperature(103) and has a larger exciton binding energy (~60 meV) than other wide band gap semiconductors.(107) ZnO is also an attractive material as it can be prepared by several techniques including spray pyrolysis(108), pulsed laser deposition(109), rf sputtering(103), chemical vapour deposition(110) and solution processed methods(111), providing good TFT performance compared to primary amorphous oxide semiconductor technologies.(3, 111) Rocksalt, zinc blende and wurtzite are possible crystal structures for ZnO(112) (113), though the rocksalt and zinc blende are generally thought of as metastable phases(114) (115). Figure 3 (a) displays the differences in the lattice configuration for different crystal structures of ZnO. Zinc blende crystal structures are stabilised through growth on substrates with a cubic lattice structure and rocksalt configurations are only stable at room temperature at much higher pressures(116) (~ 9 – 10 GPa). Consequently, wurtzite ZnO is the most commonly observed and studied crystal structure for ZnO and has a reported direct bandgap of ~ 3.3 eV at room temperature, depending on the specifics of the sample preparation method and post treatment annealing(117) (118) (119). There have been studies comparing the optical properties and bandgap of wurtzite versus rocksalt(120) and wurtzite versus zincblende(121) ZnO structures under similar environments and growth mechanisms with small differences observed between the reported bandgaps. Figures 3 (b) and (c) display an enlarged view of the wurtzite crystal structure and an example of a calculated band structure for ZnO, respectively.

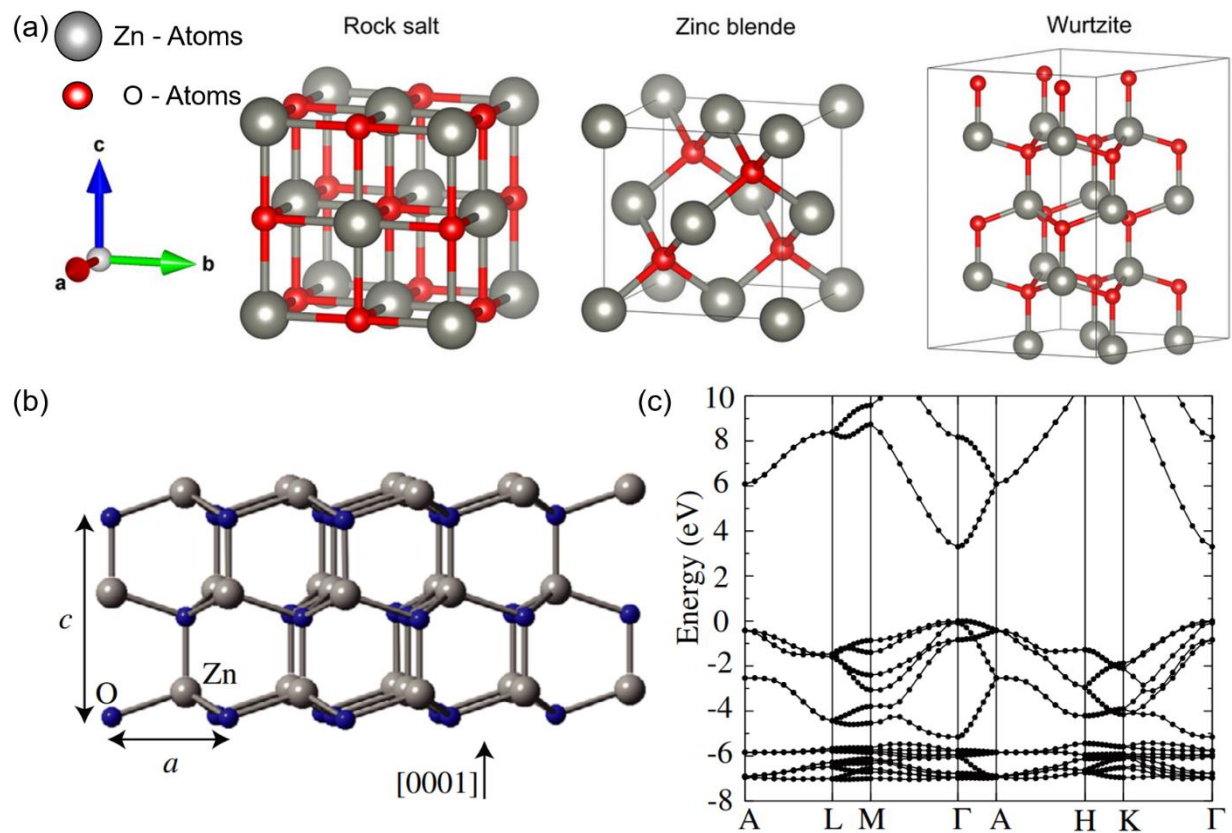

**Figure 2.** (a) Zinc and oxygen atom arrangements in common zinc oxide crystal structures(113). (b) An enlarged view of the ZnO wurtzite crystal structure(122). (c) A typical band structure diagram calculated for wurtzite ZnO showing a prominent direct band gap present(122).

ZnO is reported as being the best binary compound for oxide TFT application,(123) however, ZnO does not meet the electrical performance standards set by ITO due to a higher resistivity from a lower majority carrier concentration.(124, 125) In order to improve its electrical properties and reduce sheet resistance and resistivity, while also maintaining a low-cost, non-toxic and stable material, ZnO can be doped with trivalent electron donor metals such as Al or Ga to ensure high free-carrier mobility and transparency via the Burnstein-Moss shift that gives metal oxides their useful attributes.(126) Group 13 elements such as Al, Ga and In can be added as electron donors to the conduction band, increasing electron concentration and TFT performance.(26) Modulating the carrier concentration of a material will also affect its optical properties, specifically the plasma frequency, dielectric constant or refractive index.(127) Introducing a dopant such as Al into the crystal lattice can also produce tensile stress in the crystal lattice. These stresses can cause band-gap widening and alter the optical properties of the material.(128)

The suitability of ZnO to a large range of applications and devices is derived from high thermal conductivity, high electron mobility along with a wide and direct band gap.(129) These give ZnO a multifunctional nature and, coupled with the diverse family of nanostructures that can be formed from relatively simple, low-cost and scalable techniques, explains the rapid expansion in the amount of research completed on ZnO materials in the last two decades.(23)

**Doping in ZnO**

Despite the range of advantages that ZnO materials have in the solution deposited semiconductor industry, it does not meet the electrical performance of other doped metal oxides, such as Sn-doped $In_2O_3$ (ITO), owing to a lower majority carrier concentration.(124, 125) ITO is currently one of the most popular semiconducting materials used in the area of optoelectronics particularly in flat-panel displays due to its wide band gap (> 3 eV).(130)

However, there is a drive to replace ITO with ZnO and its doped forms such as In-doped ZnO (IZO), In-Ga-Zn-O (IGZO) and Al-doped ZnO(AZO)(4) due to the larger band gap energy of 3.3 eV. Further, the use of high-performance indium-based metal oxides is becoming less favourable due to the natural scarcity and subsequent price increase. Previous reports by the European Commission in 2014, 2017(131) and 2020(132) indicated that Indium was a critical raw material, positioning indium just outside of the top ten most critical raw materials in terms of supply risk. The most recent report from the European Commission in 2023 dropped Indium from the list of critical elements in the EU zone(133). It is important to note that this decision, to drop Indium from the list of critical raw materials, was taken through consideration of global supply and EU specific sourcing data, compared to previous reports only considering global supply. There are few ITO manufacturers based in the EU specifically, thus a lower overall demand for the EU zone(133). Globally, ITO production accounts for the majority of indium usage, with estimates of about 60% of the element being used in ITO manufacturing(133) (134). Recently, the success of indium-based photovoltaic thin films of copper-indium-gallium-selenide (CIGS) in solar cells, coupled with the future predicted demand of these forms of renewable energies for achieving greener energy production and meeting reduced $CO_2$ targets, points towards a significant increase in indium demand in the near future(134) (135) (136) (137). Figure 4 displays a forecast of indium demand if CIGS-type solar cells become dominant as solar cells in the future(135).

Of the zinc-based metal oxides mentioned previously, AZO is one that shows particular promise. The composition serves as an indium-free alternative which lowers expense due to the higher relative abundance of zinc in the Earth's crust (75 ppm for Zn opposed to 0.16 ppm for In).(124) ZnO has an intrinsic n-type conductivity which make it useful in applications such as thin film solar cells. Native, donor defects such as zinc interstitials and oxygen vacancies are reported to be responsible for the reason why ZnO can be further n-type doped by the inclusion of Group 13 elements such as Al, Ga and In.(129) Elements belonging to the boron group act as shallow donors when substitutionally doped into the ZnO crystal structure. These materials act as n-type dopants due to the presence of the loosely-bound, extra valence electron which is more easily excited into the conduction band.

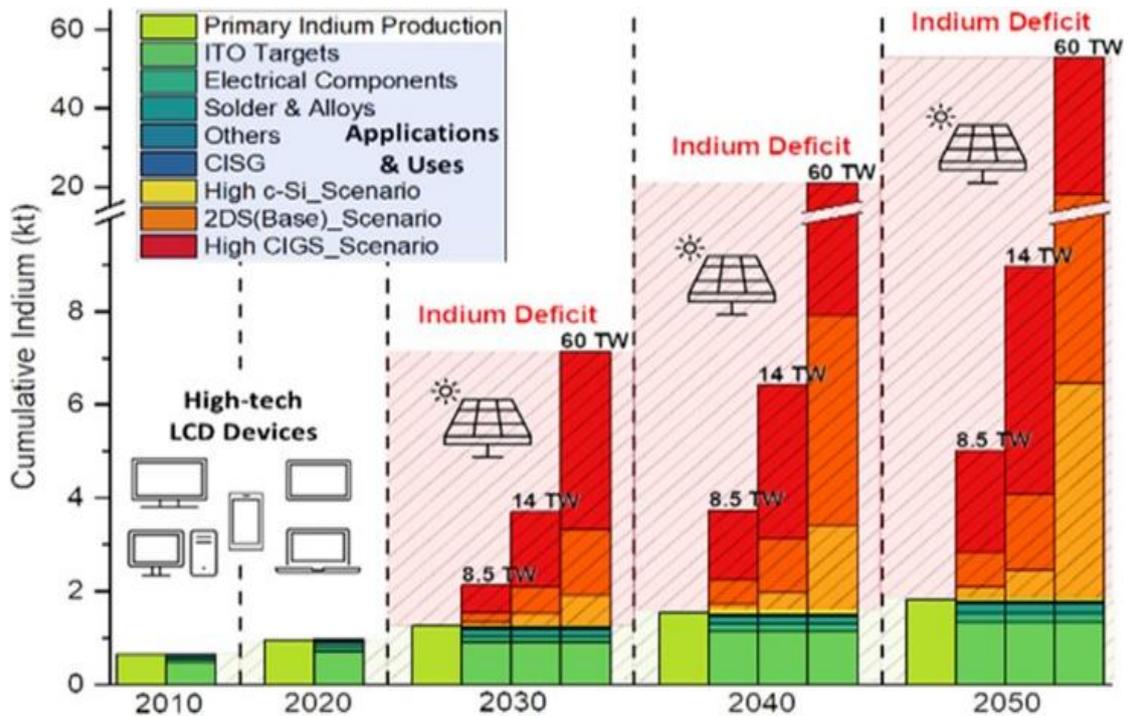

**Figure 3** A forecast of indium demand if copper-indium-gallium-selenide thin films become dominant in solar cell devices to meet green energy targets in future decades(135).

The demand for materials that are low cost and high performance drives research into TCO and amorphous oxide thin films, leading to a range of crystalline and amorphous oxides for TFTs including indium-gallium-zinc oxide (IGZO), indium zinc oxide (IZO), and zinc-tin oxide (ZTO), aluminium zinc oxide, (AZO), hafnium-zinc-tin oxide (HZTO) and others.(3, 19, 105, 123, 138) Al-doped ZnO TFTs have been grown at 120°C(66), proving that the development in unintentionally doped ZnO(125) is a feasible option for flexible transparent TFTs where high mobility is not the dominant requirement.(139)

Altering the carrier concentration of a metal oxide such as ZnO with the addition of an interstitial dopant will affect the optical properties.(127) The refractive index of the material, dielectric constant and plasma frequency are all modulated by the charge carrier concentration while the band-gap can also be widened by the stresses produced in the crystal lattice from the inclusion of a dopant atom.(128) Optical reflectivity, which is subject to change with the addition of a dopant, is of key importance to materials researched in optoelectronics, particularly in transparent TFT devices and solar cells.(75) Controlling growth, crystallization and thickness of thin films of these materials is required in order to improve and are heavily reliant on both the choice and concentration of dopant materials. This extends the impact of dopant choice to characteristic properties, primarily electronic conductivity and optical transparency, which are key factors in material suitability in optoelectronic devices.

Doping in ZnO provides beneficial electronic properties by the increase in carrier concentration through the introduction of donor-type defects and substitutional atoms with higher numbers of free electrons in the ZnO lattice. Widely used dopants of ZnO are indium, gallium, aluminium and tin as they provide favourable electronic characteristics while also allowing for the material to be grown in transparent thin film form. The first report of the inclusion of indium and gallium into the zinc oxide lattice (IGZO) and its use as an amorphous oxide semiconductor is presented by Hosono in 2004.(3) Other dopants that are presented for ZnO include hydrogen(140) (141), nitrogen(140) (142), fluorine(143) (144), titanium(145) (146), magnesium(147) (148), hafnium(149) (150) and alkali metals.(64, 70, 151). Exploring potential dopants in ZnO films is consistently a research area of interest, with many reports published on the effects of dopants on the electronic, optical and physicochemical properties bestowed by particular dopants, their concentrations or combinations thereof(152) (153) (154). Looking for indium-free alternatives to low resistance, high field-effect mobility thin film semiconductors has driven much research into the area of ZnO materials doped with readily available, low cost elements.

**Solution Processing**

Further to the improvements seen in metal oxides in their controllable electrical characteristics, ZnO and other transparent conducting oxides also show a benefit in their applicability to growth/deposition methods that avoid the high cost and low vacuum processes required for the production of e.g. amorphous silicon devices while also featuring comparatively higher electron mobility(155) (156). These are important considerations for applications such as mobile phone display screens which typically feature very high resolutions and screen refresh rates(157). Solution processing is key among these methods and offers fabrication at low temperature, high compositional control, large area uniformity, transparency, flexibility and also allows for the use of a diverse range of substrates, when compared to a-Si:H and polycrystalline Si(158) (159). A general process flow with a range of solution processing techniques are summarized in Figure 5.

Solution processes have a number of advantages over high-vacuum and photolithographic methods(4, 31) in that they allow for large-scale, cost effective fabrication and offers a high degree of compositional control(29) (75). It is crucial that high crystal quality epitaxial-like thin films can be formed from solution processing to compete with physical deposition methods. Further to the benefits gained from a solution-based approach to metal oxide deposition, these methods enable the fabrication of devices on a wide range of substrates. These, most notably, include flexible/stretchable materials to drive development into the field of wearable and plastic electronics(160). From a market perspective, wearable and flexible electronics is a rapidly growing field with forecasts anticipating a market growth of up to 20% per year up until 2028, with a predicted market valuation of about 150 billion euro in 2028(161) (162). Evidently, the demand for flexibility in wearable display screens is clearly a pressing concern to keep pace with this growing market and solution processing of metal oxide TFTs is one avenue which can be explored towards meeting this demand.

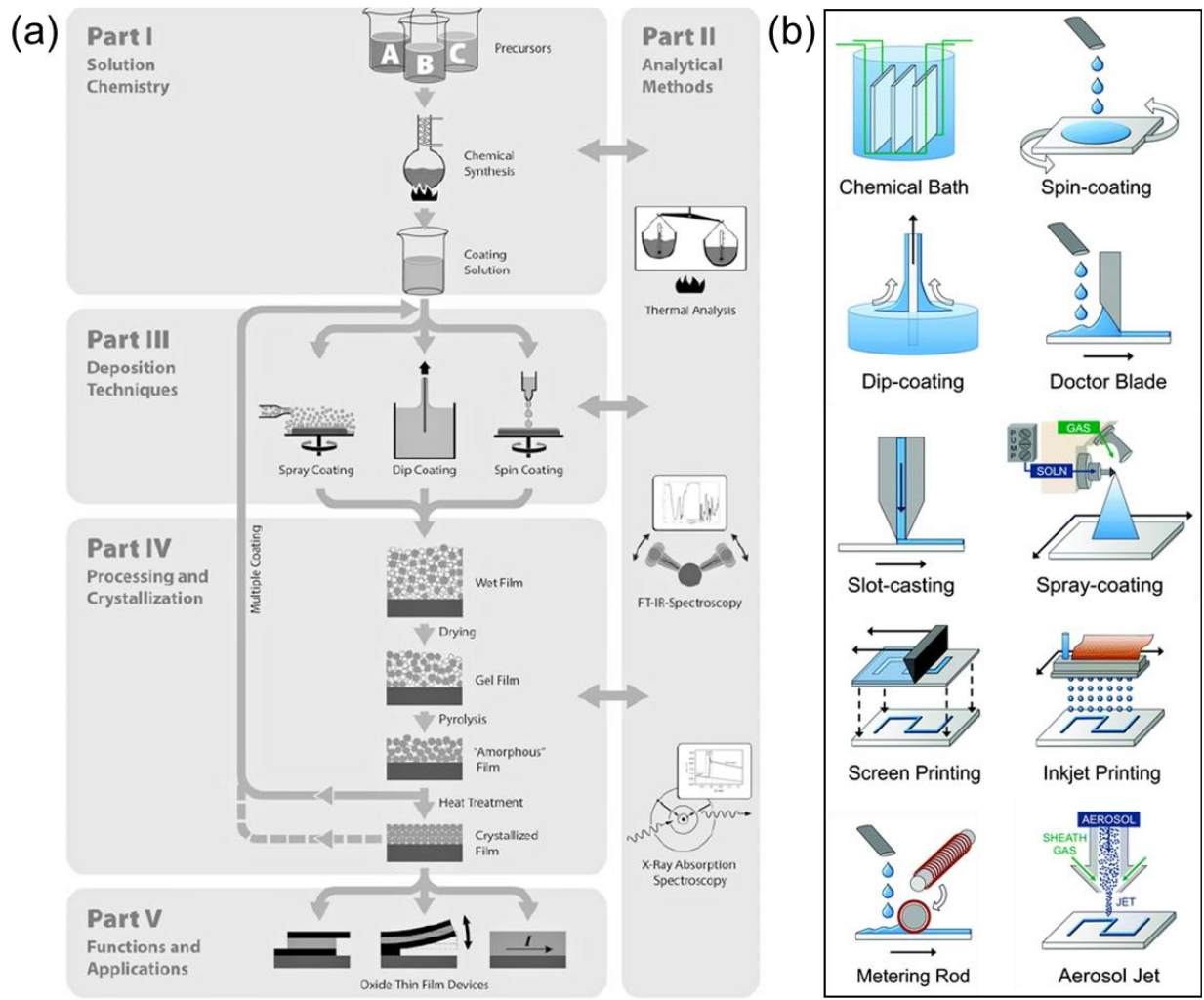

**Figure 4.** (a) General schematic for the process of solution deposition of metal oxides(163). (b) Various solution deposition methods.(31)

**Spin Coating**

Solution-based deposition methods are key for meeting the ever-growing demand for high-mobility, low-cost, semiconducting materials for use in TFTs for display technology. With the advent of flexible electronics and curved display devices, there has been a growth in the importance of industrial process scalability and the requirement for uniform coverage of non-standard substrates when developing new optoelectronic semiconductors. In general, there is a great interest in the optimization of solution processed techniques by improving our understanding of meniscus-based techniques(164) (165) (166) .

Solution processing is a term that covers a large range of growth methods, primarily including spin coating(167), dip-coating(168) and spray coating/pyrolysis(70). Among these, spin coating has received the most attention in the last decade,(29) however, there is promising alternatives in solution processing in high throughput methods such as slot casting/die-coating,(169) inkjet printing(170) and gravure/roll-to-roll printing(171). Solution processing techniques such as doctor blading quantum dot solutions(172) (173), electrohydrodynamic jet printing(174) (175) and flexographic printing(176) (177) are among some modern methods for thin film preparation. Figure 5 (b) highlights the most prominent solution processing methods while Figure 5 (a) describes the fundamental process for oxide film formation through a solution-based methodology.

In the context of the deposition of metal oxide thin films, the spin coating process involves the deposition of the sol-gel or precursor solution across the surface of the substrate, once the substrate has been mounted on the chuck of the spin-coater. To avoid streaks or swirling patterns, it is important to evenly cover the surface with precursor solution. The substrate is then spun at a desired amount of rotations per minute (RPM), which includes a desired ramp time. Films for metal oxide growth are typically spun for approximately 30 – 60 s with a 5 – 10 s ramp up time. In general, the longer and more aggressively a substrate is spun, the thinner the resulting film layer will be.

Spin coating is a popular technique and well understood, however, the physical equations governing the resulting film thickness are not frequently discussed in works presenting solution deposited semiconductor materials. Emslie et al., who presented a solution of general equations to describe the flow of fluid on a rotating disc, initiated the use of these equations to predict film thickness from a spin coating process.(178) The equation below describes the relation of instantaneous fluid height, $h$, to radial velocity, $v$:

$$v = \frac{1}{\eta}\left(-\frac{1}{2}\rho\omega^2 rz^2 + \rho\omega^2 rhz\right) \qquad (Eq.\ 1.1)$$

where $\eta$ is the fluid viscosity, $\rho$ is the fluid density, $\omega$ is the angular velocity and $r$ and $z$ belong to the circular polar coordinates $(r, \theta, z)$, $z$ being a height component from 0 to $h$. While these equations are based on multiple assumptions, including the assumption that the rotating plane is infinite in extent and does not account for non-Newtonian fluids of high viscosity, this work pioneered the investigation into the parameters affecting fluid thinning based on rotational speed and material properties.

Importantly, film thickness is not solely a result of thinning across the surface and rapid centrifugation of excess liquid off the sides of the substrate, the evaporation of the liquid and the geometry of the substrate plays a significant role in the final thickness and uniformity of the wet film(179) (180). While being held at a constant angular velocity, the combination of fluid flow and evaporation work in tandem to thin the surface liquid. Hence, choice of solvent impacts the potential layer thickness of deposited films as with an increase in precursor volatility, there is a decrease in film thickness. Rate of evaporation is a crucial factor in spin coating of thin films as it can lead to potential defects in coatings such as striations or non-uniformity in coating thickness.(163)

While spin coating presents a deposition method that is not fully material efficient, as opposed to other printer-based techniques, it remains an attractive technique in both academic and industrial semiconductor fabrication. This is heavily influenced by the low-cost, widely available and open-air compatible deposition method that is capable of generating highly uniform and controllably-thin metal oxide films. As will be examined further on, such methods are very useful when depositing iterative coatings of similar materials, or heterostructured layers in the form of solution processed superlattice materials.

**Precursors**

In the area of solution deposited semiconductor oxide materials, the choice in precursor is crucial in controlling the resulting film properties such as morphology, thickness, stability and most importantly, charge carrier mobility. The choice of precursor will impact the physical properties of the film such as density, prevalence of defects and impurities, all of which will impact the optoelectronic performance of the film(181) (182). Knowledge on the impact of precursor solution properties allows for control over key material properties of the metal oxide and also enables developments to be made on this foundation via the addition of inclusions, catalysts and dopants. The precursor concentration in ZnO films prepared via spray pyrolysis was found to influence the crystallite size and number of defects in the films(183). Similarly, optimisation of the precursor used in spray pyrolysis through the addition of a stabilizer, such as ammonium acetate, was found to reduce defects in the final film and subsequently increase the electron mobility of the ZnO film(184). The foundation of solution formation lies in the dissolution of a precursor in a suitable solvent that has physical properties that allow for deposition via the selected method e.g. a viscosity and evaporation rate that allow for dip coating of a continuous, uniform thin film.(185) Further to this, precursors should be non-reactive in storage to avoid an impact of solution decomposition on the resulting material but simultaneously be sufficiently reactive to allow for full chemical conversion under heat treatment, hydrolysis or other treatments.(186)

The liquid precursors at the centre of solution processing can be formed through a number of mediums or processes such as sol-gel mixtures and from alkoxides and carboxylates.(163) In its simplest form, solutions for metal oxide deposition can be metal salt-based precursors(187) or nanoparticle-based precursors,(188) with a preference on the metal salt approach due to improved uniformity and interface quality.(25) Precursor solutions formed at alternate temperatures or of various composition will affect the properties of the resulting thin films, which has generated the vast body of work completed on metal oxide optimisation for optoelectronic applications.

A multitude of precursor solutions are now well known for the formation of ZnO thin films, particularly in the area of ZnO as the semiconducting layer in solution deposited thin film transistor devices. A common choice of precursor materials for the formation of thin films of ZnO is zinc acetate dihydrate ($Zn(CH_3COO)_2 \cdot 2H_2O$) and ethanolamine (MEA, $NH_2CH_2CH_2OH$) dissolved in a solvent of 2-methoxyethanol ($CH_3OCH_2CH_2OH$). This composition of precursor solution is frequently seen in the use of spin coating for the deposition of doped and undoped ZnO films.(189-195) This sol-gel precursor foundation is used outside of the field of ZnO materials in TFT devices, where Lin et al. show the impact of the concentration ratio of a zinc acetate precursor on the performance of resulting ZnO films as buffer layers in organic solar cells.(196) Figure 6 shows a diagram of the mechanism of formation for ZnO films using the zinc acetate, MEA and 2-methoxyethanol approach.

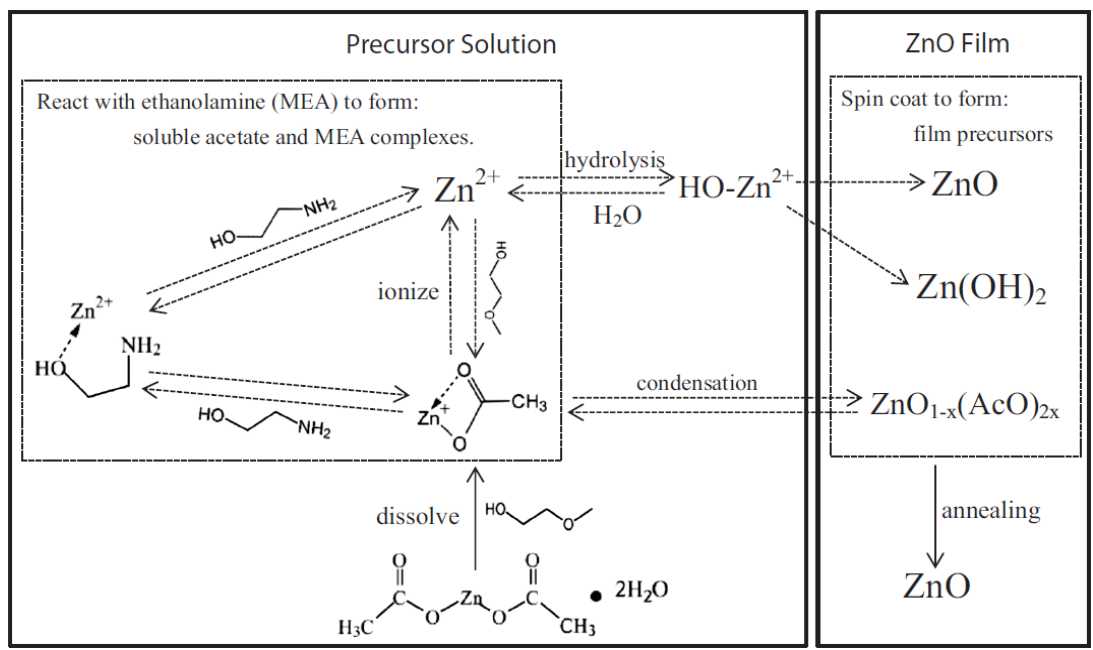

**Figure 5.** Schematic of the sol-gel reaction of zinc acetate dihydrate and ethanolamine to form ZnO films.(196)

Alternatively, TFT devices involving ZnO doped and undoped materials formed through different precursor materials have been reported, with a variety of resulting field-effect and saturation mobilities. Park et al.(151) report an approach of dissolving pure zinc oxide powder in ammonium hydroxide, along with the addition of alkali metal dopants, to fabricate devices which produced a field effect mobility of 11.45 $cm^2\,V^{-1}\,s^{-1}$. Liu et al.(197) notably produced indium zinc oxide (IZO) TFT devices through a precursor formed from dissolution of zinc nitrate in deionised water, resulting in binary metal oxide formation that produced devices with reported field effect mobility of 2.79 $cm^2\,V^{-1}\,s^{-1}$. Pujar et al.(198) report formation of indium zinc tin oxide (IZTO) devices with saturation mobility of 4.21 $cm^2\,V^{-1}\,s^{-1}$ from dissolution of zinc nitrate in 2-methoxyethanol. In solution processing of metal oxides, the addition of dopants is further enabled and controlled through the addition of the dopant precursor to the main precursor solution and mixing prior to deposition(199).

**ZnO Materials by Solution Processing**

Reviews of metal oxide materials with a focus on solution processing present results that put the deposition of thin films at the foreground.(4, 23, 31) This is due to the primary drive for development of new metal oxide materials and the refinement of processes for existing materials being the improvement in device performance of thin film transistors, for which the requirement is continuous, uniform thin films.

However, there is an abundance of other structures being formed from ZnO via numerous deposition methods encompassing physical deposition techniques and wet chemical approaches. As per the scope of this review, a primary focus on the formation of ZnO structures, outside of thin films, through solution deposition methods is presented.

Xu et al.(200) and Ko et al.(201) present arrays of nanowires grown via chemical approaches and can be seen in Figure 7. Xu et al. present a seedless chemical approach that shows substrate-perpendicular growth of uniform ZnO nanowires with an aim toward application in FETs and nanopiezoelectronics. Using a hydrothermal method of zinc nitrate in water, Ko et al. report less ordered arrays of nanowire networks that give an assumption-based field effect mobility of 0.2 $cm^2\,V^{-1}\,s^{-1}$.

ZnO nanotubes with orientation perpendicular to the substrate are grown from ZnO seeds prepared via the sol-gel method by Chu et al.(202) and then developed through evolution from resulting nanorods into arrays of ZnO nanotubes via a chemical bath solution. The resulting structures are controlled by adjustment of pH levels, temperature and growth time.

Growth of ZnO nanorods from solution methods was also presented by Sun et al.(203) and Yi et al.(204) Yi et al. explore the potential application of these nanorods, which include switching devices, field effect transistors, UV detectors, diodes and sensors. The focus of the work presented by Sun et al. in 2005 is the fabrication of TFT devices from spin coated arrays that present mobility values of 0.20 cm$^2$ V$^{-1}$ s$^{-1}$. More recently, in 2017, Kumar et al.(205) report field effect mobilities of 9.04 cm$^2$ V$^{-1}$ s$^{-1}$ for laterally aligned ZnO nanorods, which also serves to highlight the drive and advancement toward higher mobilities over the last decade.

A structure which is perhaps furthest from the scope of this work while still being a solution processed zinc-based metal oxide is the tetrapod, shown in Figure 7. A thorough review of the formation of ZnO tetrapods is presented recently by Mishra et al.(206) This review highlights tetrapod structures that are grown via numerous methods, including hydrothermal and wet chemical solution-based techniques and their potential applications.

Cho et al.(207) present a technique for the formation of ZnO nanoparticles using zinc acetate dihydrate dissolved in methanol and the subsequent preparation of multiple aqueous ZnO inks using various concentrations of nanoparticles. TFTs fabricated from this method report field effect mobilities of 1.75 cm$^2$V$^{-1}$s$^{-1}$, supporting the motivation for further research into solution-based ZnO materials for flexible printed electronics.

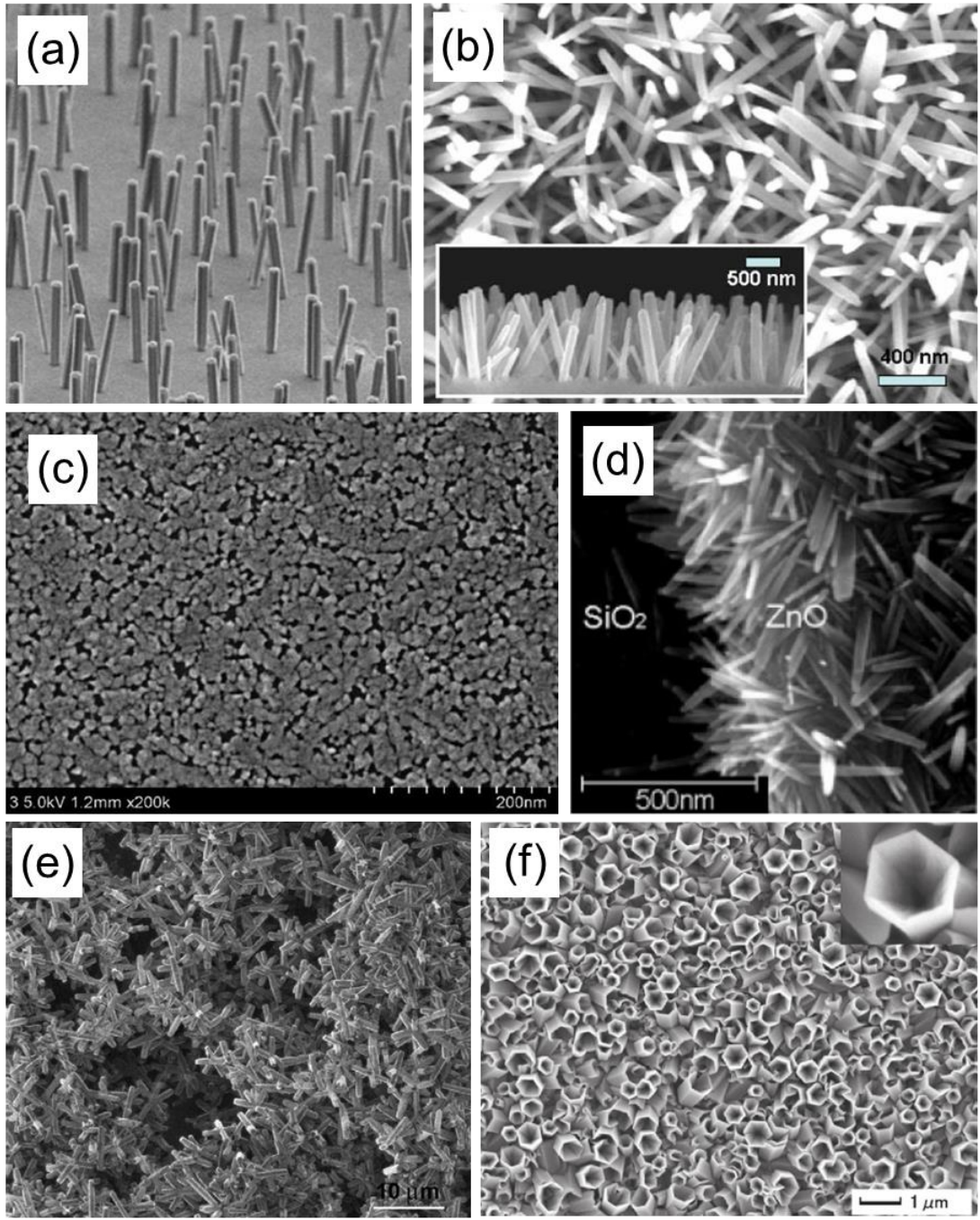

**Figure 6.** SEM images of representative ZnO structures grown from solution-based methods. (a) Nanowires grown on Si substrates from a nutrient solution.(200) (b) Dense array of hydrothermally grown nanowires.(201) (c) Surface SEM of ZnO nanoparticles from aqueous ink solution.(207) (d) Interface of $SiO_2$ and ZnO nanorod film, formed via hydrothermal growth of nanocrystals.(203) (e) Architecture consisting of hydrothermally grown ZnO tetrapods.(208) (f) ZnO nanotubes resulting from etching of nanorod arrays in aqueous solution.(202)

Naturally, due to the breadth of the research conducted using ZnO semiconductor materials, the examples above represent only a limited view into the full area of ZnO nanostructures via chemical methods. An important structure related to metal oxide growth for use in modern electronic devices is the superlattice, an amalgam of iteratively deposited layers. This structure is discussed in detail below.

**Quasi-Superlattice Structures**

At oxide interfaces with the dielectric, between homo- or hetero-phase interfaces in superlattices, trapped charges in an oxide layer or interface from electric field-induced removal of majority carriers leave a net positive charge density on the interface. These trap sites often cause shifts in threshold voltage and affect leakage characteristics.(209) There is a focus in the area of superlattices on the generation of heterojunctions due to the enhancement of charge carrier mobility and the resulting impact to the metal oxide performance in thin film transistor devices based on interfacial properties(210). Some works have focus on developing solution-processed homojunction materials, where double-stacked semiconductors of the same material are deposited, with reported improvements to recorded field-effect mobility(211) (212). Although some quasi-superlattice structures comprised of hetero-phases provide higher mobility at the interface compared to the corresponding bulk material layer, homogeneous QSL structure have received limited attention. Such structures may provide controllable thickness and interfacial transport characteristics, but also a consistent refractive index and transparency compared to hetero-SL structures.

There has been much work completed in the last decade on the fabrication and characterization of superlattice structures. A significant contribution to this field of knowledge has been generated by Anthopoulos et al. on the area of quasi-superlattice structures developed from a solution method.(64-67, 213, 214) Quasi-superlattices have been investigated for their potential to provide improved electrical characteristics through 2D-transport properties. Lin et al. show solution processed heterojunctions in a quasi-superlattice configuration that suppose the formation of 2D electron gas (2DEG) near the heterointerfaces.(214) Devices in this work are fabricated using $In_2O_3$, ZnO and $Ga_2O_3$ in various layered structures. Here it is reported that for metal oxides deposited as superlattices, the performance level of the fabricated TFT is not dependent on the bulk properties or mobility of the individual semiconductors.

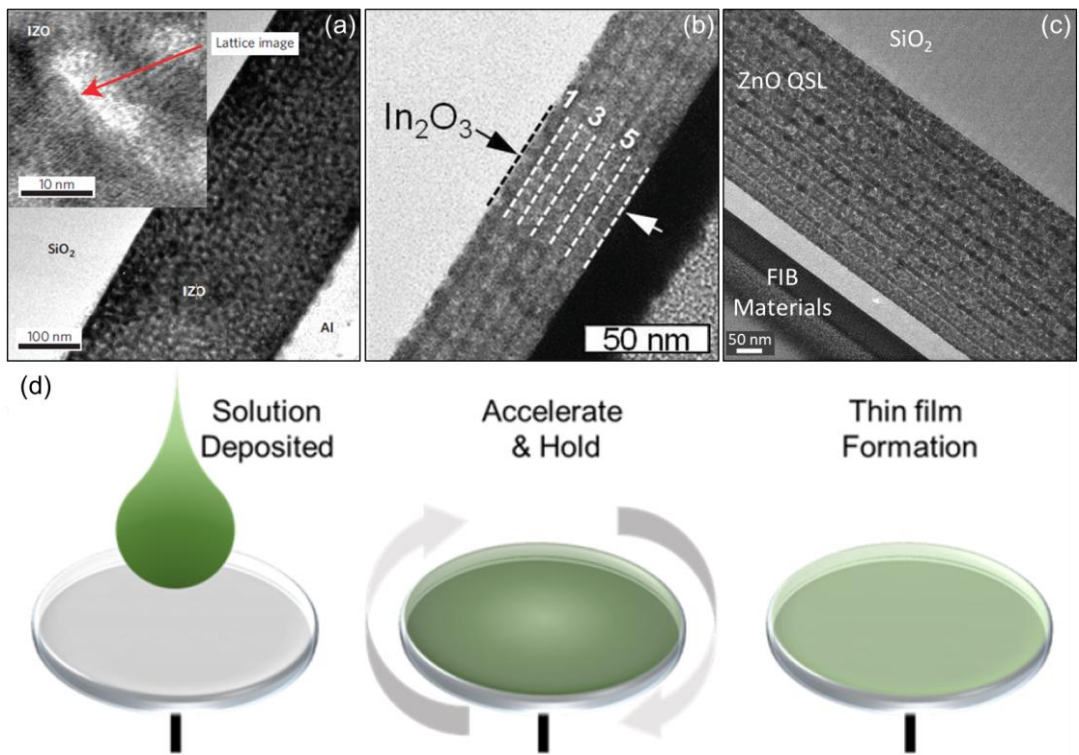

**Figure 7**. QSL structures fabricated using iterative spin coating deposition techniques, reported by (a) Banger et al.(19) (b) Labram et al.(67) and (c) QSL structure of ZnO reported in this thesis.(167) (d) Schematic of the spin coating process for deposition of thin films.(215)

At the heterointerfaces within the QSL there is a supposed 2DEG formed that separates the charge carriers from the donor/acceptor sites which leads to less charge carrier scattering and higher mobilities. A 2DEG formed at the interface has also been reported using ZnO/ZnMgO interfaces.(216) Resulting mobilities of $> 40$ $cm^2$ $V^{-1}$ $s^{-1}$ highlights that devices fabricated from QSLs show great potential in improving device performance based on design engineering.(214) Faber et al.(65) furthered this research by showing solution-grown heterojunctions of $In_2O_3$ which exhibit peak mobilities of $> 45$ $cm^2V^{-1}s^{-1}$ that rely on seamless interfaces of crystalline layers grown from solution deposition methods to increase electron transport properties. These works from Anthopoulos et al. show that the properties of electrical devices such as TFTs(214) and diodes(67) benefit from band structure engineering, through the employment of alternating layer deposition and QSL formation. Banger et al. produce a superlattice structure, as shown in Figure 8(a), using their well-known "sol-gel on a chip"(19) method that shows a quasi-superlattice structure composed of indium zinc oxide. Homointerfaces of IZO are displayed where there is clear definition between iteratively-deposited layers, however, the impact on the electrical properties is not discussed. Sequential deposition of $In_2O_3$ is presented by Faber et al. with a motivation to demonstrate a control of ultrathin layer dimensionality and conduction band characteristics from solution at relatively low temperatures (200 °C in air).(65)

Fabrication of layered metal oxide TFTs is not limited to just bilayer structures of alternating materials; trilayer and multilayer TFTs are also frequently explored in the literature. The Anthopoulos group(214) explored the electron mobility of different types of metal oxide layered TFTs, namely heterojunctions versus quasi-superlattices, with various numbers of layered materials, all deposited via spin casting. Of the TFTs constructed of ZnO, $In_2O_3$ and $Ga_2O_3$ metal oxide layers, the trilayer quasi-superlattice structure of $In_2O_3$, $Ga_2O_3$ and ZnO featured the highest room temperature electronic mobilities calculated in the saturation regime(214). Figure 9 (a) illustrates the layered structures tested and the performances recorded. A follow-up study with these types of heterojunction structures of ZnO, $In_2O_3$ and $Ga_2O_3$ combined with a $Cs_2CO_3$ work-function modification layer for organic light-emitting field transistors, reported large area uniform light emission and significantly enhanced external quantum efficiency(217).

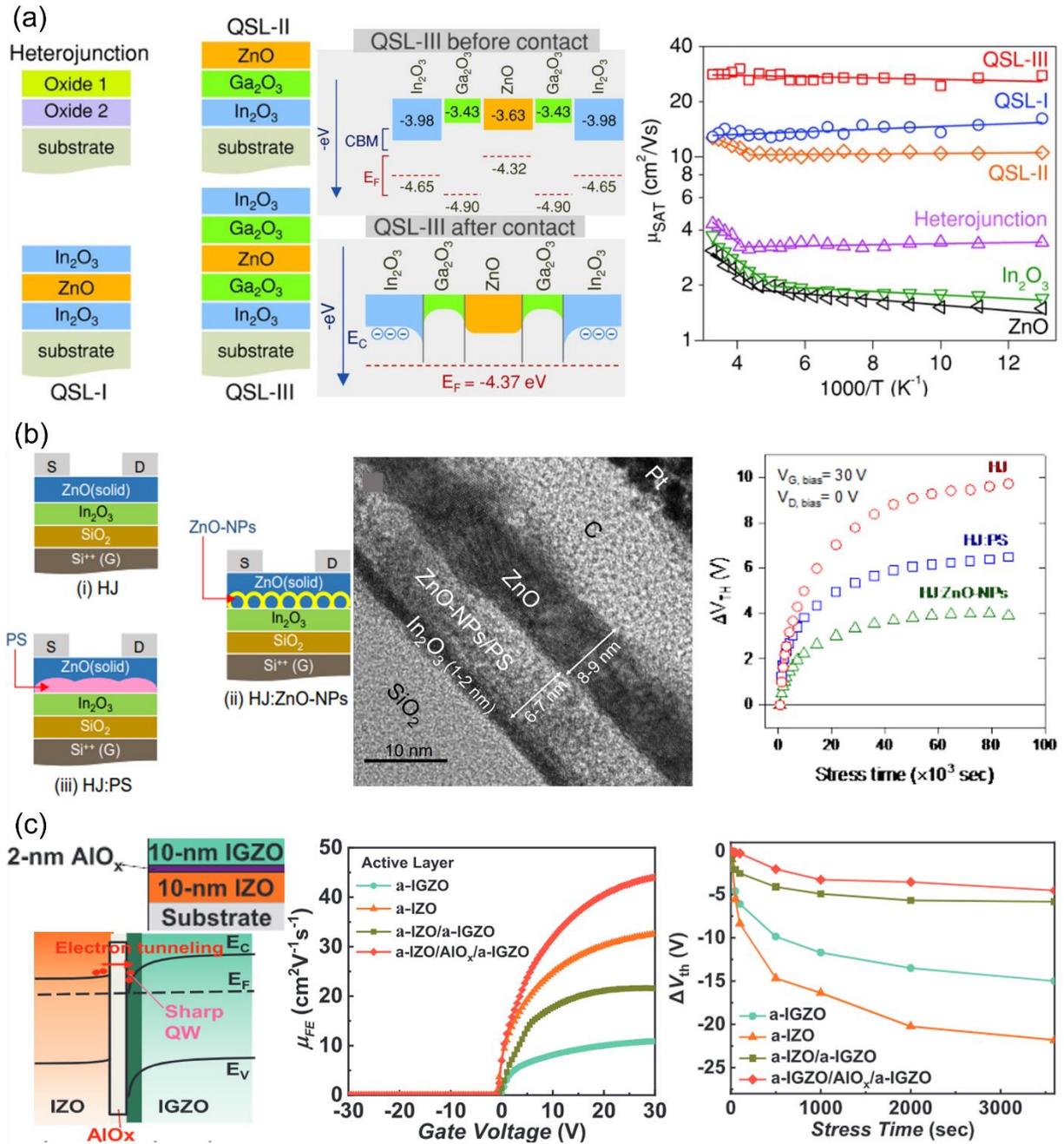

**Figure 9**. (a) Heterojunction and quasi-superlattice structures constructed from combinations of $In_2O_3$, $Ga_2O_3$ and ZnO layers, showing bandgap modification schematics and measured saturation electronic mobilities(214). (b) Various heterojunction structures of $In_2O_3$, solid ZnO, ZnO nanoparticles and ozone-treated polystyrene the lowest shift in threshold voltage shown for the structure incorporating ZnO particles(63). (c) Comparisons of electron field mobility performance and threshold voltage shift between amorphous IZO, IGZO monolayers, bilayers of IZO and IGZO and trilayers of IZO, $AlO_x$ and IGZO with strongest performance indicators for the trilayer structure(218).

That group also investigated operational stability in multilayered TFTs comprising In$_2$O$_3$, ZnO, ozone-treated polystyrene and ZnO nanoparticles by measuring changes to the threshold voltage ($\Delta V_{TH}$) under bias-stress measurements(63). The multilayered heterojunction structure displayed the smallest change to threshold voltage, indicating superior operational stability, and an associated high electron mobility, attributed to a trap passivating effect of the ozone-treated polystyrene layer(63). Figure 9 (b) displays these results alongside a schematic of the mulitlayered TFT structure. Recently, Wang et al.(218) explored the performance of sputter coated multilayer TFTs comprising various configurations of InZnO, InGaZnO and AlO$_x$ layers. The trilayer structure consisting of a thin (2 nm) layer of AlO$_x$ sandwiched between thicker InZnO and InGaZnO layers, featured the best performance in terms of highest field effect mobility and highest stability with the smallest voltage shift recorded under negative bias illumination stress(218). Figure 9 (c) shows the device layer configuration and results recorded for these materials.

Aside from multi-layered heterojunction structures, multi-layered homojunctions and homogenous quasi-superlattice structures have also been explored in the literature, typically in the form of iterative layers of the same metal oxide material deposited through solution-processed techniques. From investigations into iterative spin-coated ZnO and Al-doped ZnO layers in an homogenous quasi-superlattice arrangement within our own group(75), iterative coatings of ZnO were found to influence the crystal growth orientation of the ZnO material, as measured through X-ray diffraction measurements. Single layers of ZnO displayed only m-plane growth orientations, while 20 layer homogenous quasi-superlattices displayed a dominant c-plane growth orientation. Both ZnO and Al-doped ZnO quasi-superlattice structures also displayed tuneable and angle dependent reflectivity properties compared to their monolayer counterparts, for applications in anti-reflection coatings. These effects can be seen summarized in Figure 10 (a).

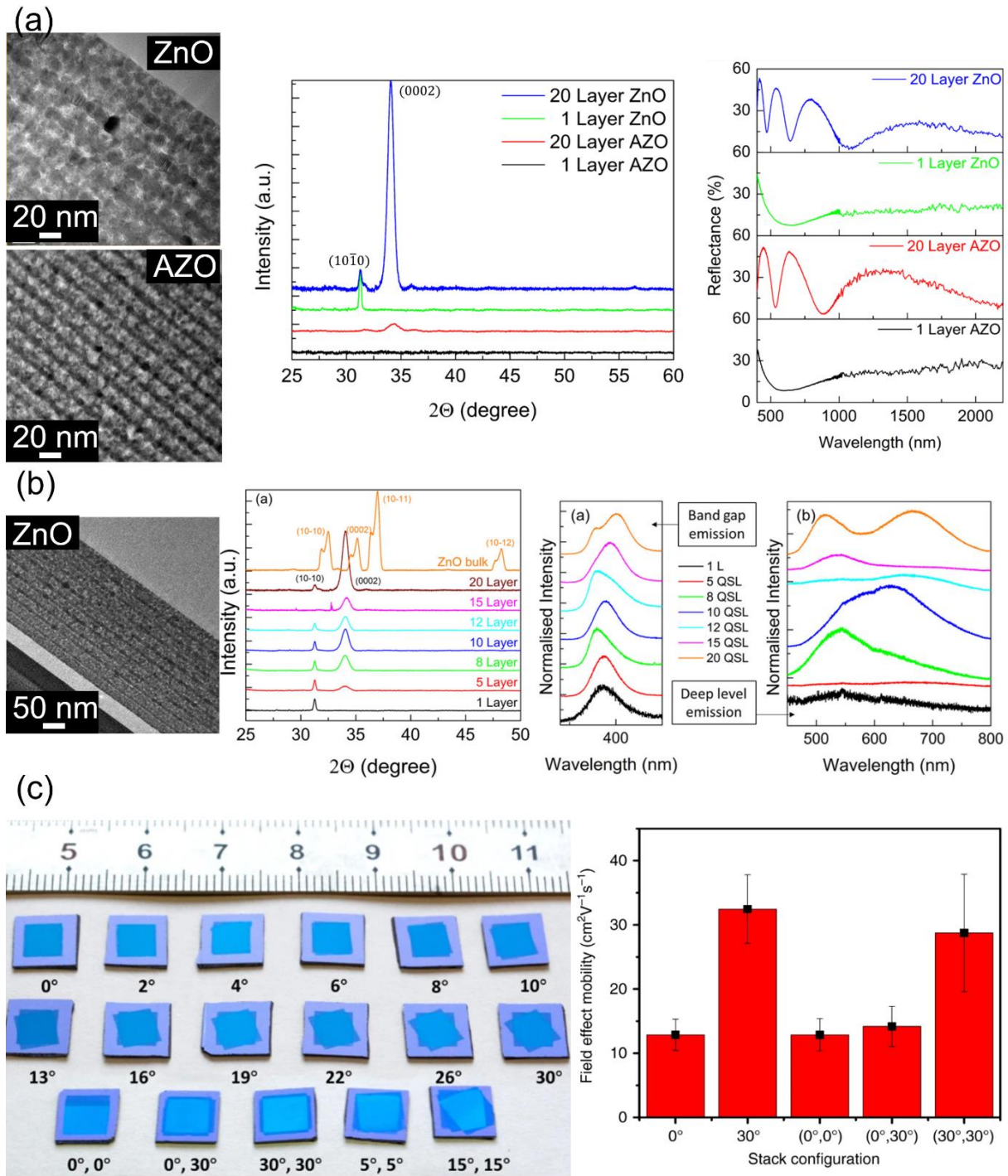

**Figure 10.** (a) Homogenous quasi-superlattice structures comprising 20 layers of ZnO and AZO, with XRD data showing a change in crystal growth direction with increasing number of layers and changes in optical reflection data for samples oriented at 45° incidence(75). (b) ZnO homogenous quasi-superlattice structures, pictured with a 10 layer cross-sectional TEM image, with XRD data illustrating the gradual change in crystal direction with number of layers and photoluminescence data showing the prevalence of sub-bandgap radiative recombination(167). (c) Various configurations of $MoS_2$ homojunctions with different interlayer twist angles for bilayer and trilayer structures. Measured field effect mobility values are also shown for selected $MoS_2$ stack configurations(219).

A follow-up study(167) found that the electronic defects in homogenous ZnO quasi-superlattice structures were directly related to the number of layers comprising the layered material. Photoluminescence spectra showed that the sub-bandgap radiative recombination via electronic defects was highest for materials with the greatest number of layers, suggesting a tuneable and predictable electronic response of layered ZnO quasi-superlattice structures(167). Figure 10 (b) displays the structure of the layered material and the tuneable photoluminescence observed from various multilayers. In a slightly different approach, multi-layered homojunctions of $MoS_2$ were developed via chemical vapour deposition(219), whereby the angle of deposition of the layers of material was the variable in structural composition. The interlayer twist angles between the layers of the $MoS_2$ homostructure were found to have direct effects on the field effect mobility recorded for assembled devices. Various different angle configurations were tested for both bilayer and trilayer homostructures with the bilayer structure with a 30° twist performing best, followed closest by the trilayer structure with layers deposited at 30°(219). These effects are illustrated in Figure 10 (c). Stacking of homogenous multilayers of $MoS_2$ with a variable twist angle is a topic of interest due to the various Moiré superlattice patterns which arise and their influence on localised electronic states and the overall electronic structure of the material(220).

These representations, grown via a myriad of different techniques ranging from chemical vapour deposition, sputter-coating and numerous instances of spin-casting or spin-coating, demonstrate that well-defined layers of metal oxides can be deposited from solution that offer controllable thickness and definition on par with other vacuum-based techniques, such as ALD.(221)

**Thin Film Devices**

As discussed earlier, metal oxide thin films fabricated from solution or other physical deposition methods have applications across many technological fields. This review is focused primarily on the applications of metal (zinc) oxide thin films and their application in the electronic and optoelectronic industries. The allure of ZnO-based thin films for use in these areas is the flexibility provided by this low cost and diversifiable material, where it can play a variety of roles in the same type of optoelectronic devices, whether as the photoresponsive semiconductor in photodetector devices, the active channel or the dielectric layer in thin film transistors, the transparent conductive electrode in solar cells or as an anti-reflective coating on a number of types of optoelectronics. This section will describe the uses of ZnO in the fabrication of modern TFTs and optoelectronics.

**Electronic Devices**

Thin-film transistors (TFTs) are the fundamental building blocks for macroelectronics(3). They are included in a variety of applications but have their largest application in display technology,(2, 222, 223) where these devices allow for the switching of pixel displays and active matrix organic LEDs (AMOLEDS) without the need for moving parts.

The typical thin film transistor employs a semiconductor material, contacted by a source and a drain electrode. This material is deposited onto an insulating dielectric layer which separates the semiconducting layer from the substrate. The substrate can and is often used as gate electrode for the TFT device. A potential bias is applied across the source drain electrodes. In "always-off" devices which are dependent on the semiconducting material chosen, this would result in no passing current as there is an insufficient electric field to move the charge carriers throughout the material. Applying a voltage between the source electrode and the gate electrode, which can be designed as a heavily doped substrate, or as a surface electrode alongside the source and drain (a process called top-gating), applies an electrical field transverse to the source-drain electrodes. This applies a bias to the charge carriers (either positively or negatively depending on the direction of the bias) and moves them to either the semiconductor surface or down to the semiconductor-insulator interface. Figure 11 (a-f) displays the most common device configurations of TFTs and the characteristic transfer and output *I-V* curves used to calculate figures of merit for metal oxide semiconductor channel materials.

High field-effect mobility TFTs were initially manufactured from low hydrogenated amorphous Si (a-Si:H),(223) however, these require vacuum-deposition processes and high fabrication costs which motivated research into cost effective, atmospheric process fabrication of alternative TFT channel materials. Within research into this area it has been shown that amorphous metal oxide TFTs can "flow high-density current ~100 times larger than that of conventional amorphous Si (a-Si) TFTs".(2) The polycrystalline-Si TFT technology currently used is also difficult to implement on lightweight and flexible plastics because of high process temperatures.(223) TFTs fabricated from organic semiconductor materials offer another cost effective alternative to a-Si devices, however, these suffer from the requirement of low work function electrodes to facilitate electron transport to and from organic semiconductors.(224) TFTs are used for the switching of individual pixels in modern displays and are a driving force for the research into conductive, transparent and flexible materials in an effort to construct devices with matching properties(225) (226). The first step in designing TFTs for use in future applications is to assess and optimise the electrical properties of the semiconductor(227) (228). The field-effect mobility, $\mu_{eff}$, is used as one of the main figures of merit for a materials suitability in TFT devices. This describes the relationship between the charge carrier mobility in the material with respect to the applied electric field.

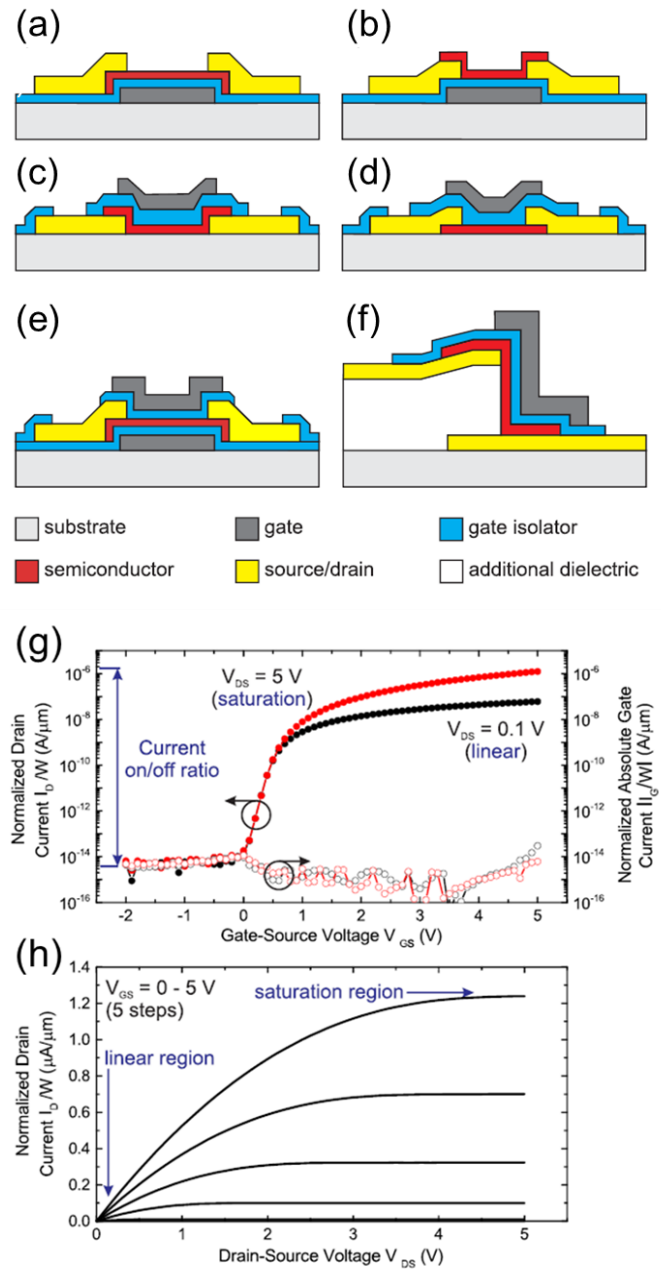

**Figure 11.** (a-f) Common device configurations for TFTs (a) bottom-gate staggered (b) bottom-gate coplanar (c) top-gate staggered (d) top-gate coplanar (e) double-gate and (f) vertical TFTS. (g) Transfer and (h) output *I-V* curves for typical metal oxide TFTs.(24)

Transparent TFTs are crucial for next generation display and interactive technologies, but are severely limited by expensive production methods(223, 229) and by a limited range of new materials that are processable outside of industrial settings. Among the TFTs, zinc oxide based materials have attracted increasing attention for use in flexible displays because of their higher mobility and lower processing temperature than conventional hydrogenated amorphous

Si TFTs.(230) ZnO TFTs have been fabricated on flexible substrates and are suited to this due to their lower processing temperature.(231) For use in TFT technology, a highly stable, high mobility, controllably thin, uniform and transparent metal oxide is required.(215)

Faber et al. have, in recent years, presented thin-film transistors with high mobilities that challenge similar devices fabricated using vacuum processes.(65) These devices are formed from a combination of spray pyrolysis and spin coating to form heterojunctions of $In_2O_3$ and ZnO at temperatures less than 250 °C that display electron mobilities that exceed 45 $cm^2V^{-1}s^{-1}$, surpassing performance of devices fabricated from single layers of either individual oxide. Their work highlights that solution methods can challenge the state of the art in vacuum-based deposition methods, leveraging the design of transistor channel materials to include synergistic material properties such as the increased charge transport mechanism presented by Anthopoulos' group.

In recent work towards the advancement of metal oxide electronics, which is in the fabrication of stable, transparent semiconductors, Wahila et al. have shown tin-doped zinc oxide films that demonstrate promising electrical and optical properties for an indium-free transparent conducting semiconductor.(232) Experimentation to produce high performance, transparent TFTs is continually ongoing and showing promising results of late. Transparent thin film transistors fabricated from undoped ZnO nanorods formed via hydrothermal methods display promising field effect mobilities of 3.86 $cm^2 V^{-1} s^{-1}$ have been presented in work from Zhang.(233) Zeaumalt et al. have demonstrated considerably increased mobilities from solution-grown ZnO TFTs via the use of a spin coated $ZrO_2$ dielectric layer.(234) By implementing a high-*k* dielectric layer in place of the commonly used, thermally-grown $SiO_2$, linear mobilities of 20 $cm^2V^{-1}s^{-1}$ were reported, an important result for achieving high field effect mobility for undoped, transparent ZnO materials.

**Optoelectronic Devices**

Investigation into the impact that dopant choice, along with their elemental ratio within the oxide film, has on the optoelectronic properties of promising or established semiconducting materials is crucial for the acceleration of solution processed oxide optoelectronic materials development. As mentioned earlier, Wahila et al. recently published a comprehensive study on the impact of Sn/Zn ratio on the optical and electrical properties of amorphous zinc-tin oxide (a-ZTO).(232) In this work, where some key results are shown in Figure 12, it is revealed that tin-poor compositions unexpectedly provide enhanced conduction properties while showing control over optical transparency with dopant concentration.

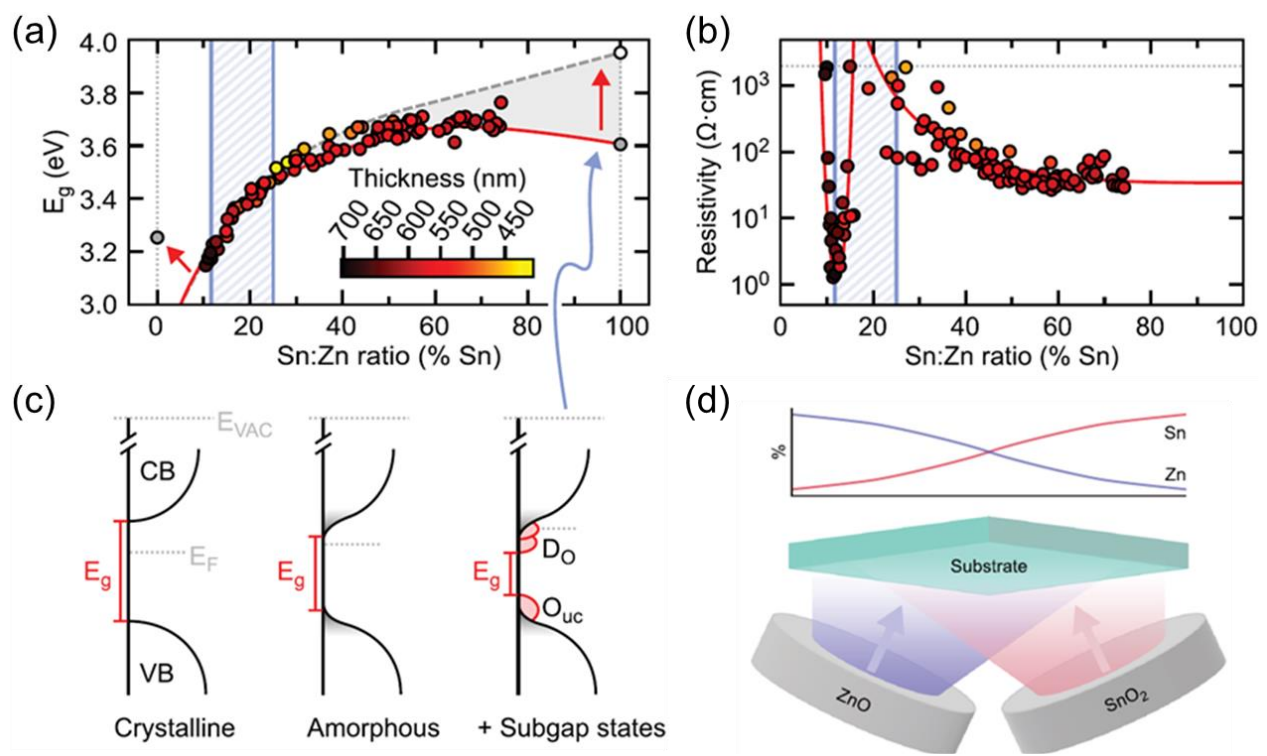

**Figure 12.** (a) Resulting optical band gap of a-ZTO thin films with respect to Sn:Zn ratio. Link with (c) shows that reduction in optical transparency through subgap state absorption occurs at tin-rich compositions. (b) Resistivity behaviour of a-ZTO films with increasing Sn:Zn ratio, highlighting that Sn-poor composition show the highest conduction performance. (c) Impact of material crystallinity and the formation of defect associated subgap states on material band gap. (d) Schematic of the co-deposition method highlighting the compositional Zn/Sn gradient across the deposited film area.(232)

Similar work in doped ZnO materials for optoelectronic device applications has been reported in recent years by Morales-Masis et al., where ZTO films are reported to have improved performance as OLED anodes over indium-tin oxide (ITO), the industry standard.(235) Films with varying ratios of Sn:Zn were fabricated via deposition by RF sputtering on glass and displayed high max electron mobilities (21 $cm^2$ $V^{-1}$ $s^{-1}$) and excellent transparency in the visible region (< 5%). This work is a key representation of the potential for improvement in device performance through the use of indium-free, low-cost and innocuous materials. Continued research into metal oxide optoelectronic materials has shown that iteratively higher device performance is achievable through adaptation of oxide material composition, deposition conditions and processing techniques.

**ZnO Photodetectors**

ZnO materials are commonly used for the detection of UV light(236) (237) (238). This is due to the nature of the wide band gap of ZnO (3.2 eV) and its doped counterpart materials that relates to a wavelength of 380 nm. This makes ZnO materials ideally suited to the detection of this, and higher energy/shorter wavelength forms of radiation. Additionally, the intrinsically low cost, high chemical stability and strong radiation hardness are all features which make ZnO attractive as a candidate for UV photodetectors(239).

This detection mechanism in ZnO-based photodetectors is based off of band gap excitation, the promotion of conduction electrons to the conduction band from the valence band by being provided with sufficient energy from the incident radiation. However, what is more rarely reported is sub-band gap excitation of ZnO materials,(240) in particular for solution deposited ZnO thin films. This allows for use of the materials in the detection of light with energy below the band gap of the material.

Early work in solution processed ZnO UV photodetectors was carried out by Basak (2003) and Xu (2006) that presented characterization and benchmarking of the performance of early sol-gel ZnO optoelectronic materials. Xu et al. demonstrate the photoconductive abilities of ZnO when doped with Al at 5%.(241) Thin films of AZO are fabricated via spin coating of zinc acetate and aluminium nitrate in 2-methoxyethanol and monoethanolamine on Si substrates. Resulting films contacted with Au show ohmic behaviour under light and dark conditions. A shift in the band gap wavelength of AZO films is seen through photocurrent and photoluminescence spectra towards a wavelength closer to 350 nm, as opposed to 370 nm for undoped ZnO, indicating suitability of doped ZnO photodetectors for excitation wavelengths below the band gap wavelength of undoped ZnO. Similarly, UV detectors are fabricated by Basak et al. via the dip coating of undoped ZnO thin films from a zinc acetate precursor solution.(242) Au-contacted films indicated the promise of solution processed ZnO for photodetection devices through slow photoresponse decay and a responsivity of 0.04 $AW^{-1}$.

More recently, Jin et al. presented ZnO photodetectors with enhanced performance through the inclusion of Au nanoparticles (NPs).(243) Devices were fabricated through multi-step spin coating protocol from a Zn ammonium complex, followed by spin coating of Au NP solutions and finally thermal evaporation of patterned Al contacts. While having the added benefits of being flexible and a high (> 90%) transmittance in the visible region, these UV photodetectors showed excellent responsivity in the UV region of up to $1.51 \times 10^5$ $AW^{-1}$ at an applied bias of 50 V. Similarly, flexible ZnO thin films grown on polyimide substrates via successive ionic layer adsorption and reaction (SILAR) were modified by Ag nanoparticles and exhibited an enhanced performance compared to films without Ag nanoparticles(244). This comprehensive characterisation of ZnO thin films for photodetection applications exemplifies, in comparison with initial work in the area, the enhanced device performance that can be achieved through continued research in ZnO optoelectronics.

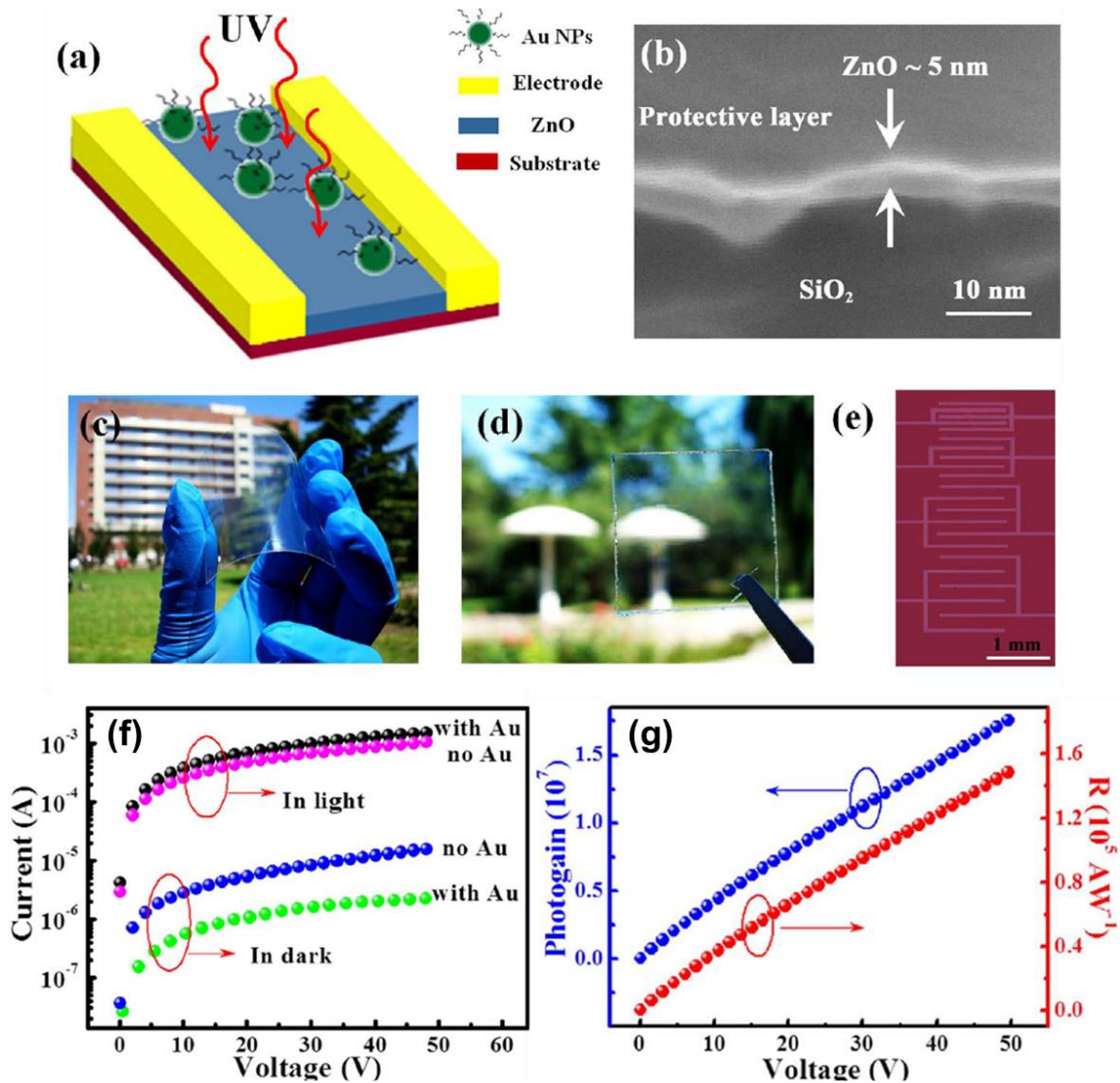

**Figure 13.** ZnO UV photodetector devices fabricated by Jin et al. (a) Device schematic and (b) cross-section showing single layer ultrathin ZnO. (c) Spin coated device highlighting the flexibility and (d) high visible-region transparency. (e) Patterned ITO electrodes deposited onto the photoconducting device. (f) *I-V* performance of devices under 350 nm illumination. (g) Photogain and responsivity of photodetecting devices.(243)

While a significant fraction of the work in ZnO UV photodetectors carried out in the last two decades is focused on ZnO nanowires, important work in 2012 from Guo et al. demonstrated solution processed ZnO nanoparticles in a hybrid photodetector that greatly outperformed inorganic semiconductor photodetectors.(245) This work utilizes a hydrolysis method to grow ZnO nanoparticles that act as the ultraviolet absorber and subsequently blended with semiconducting polymers. Similarly, Yuan et al. constructed a ZnO-based UV photodetector with a stable, rapid and repeatable response towards 365 nm UV light using 5.5 nm ZnO nanoparticles which were spin-coated into a thin film on an FTO substrate(246).

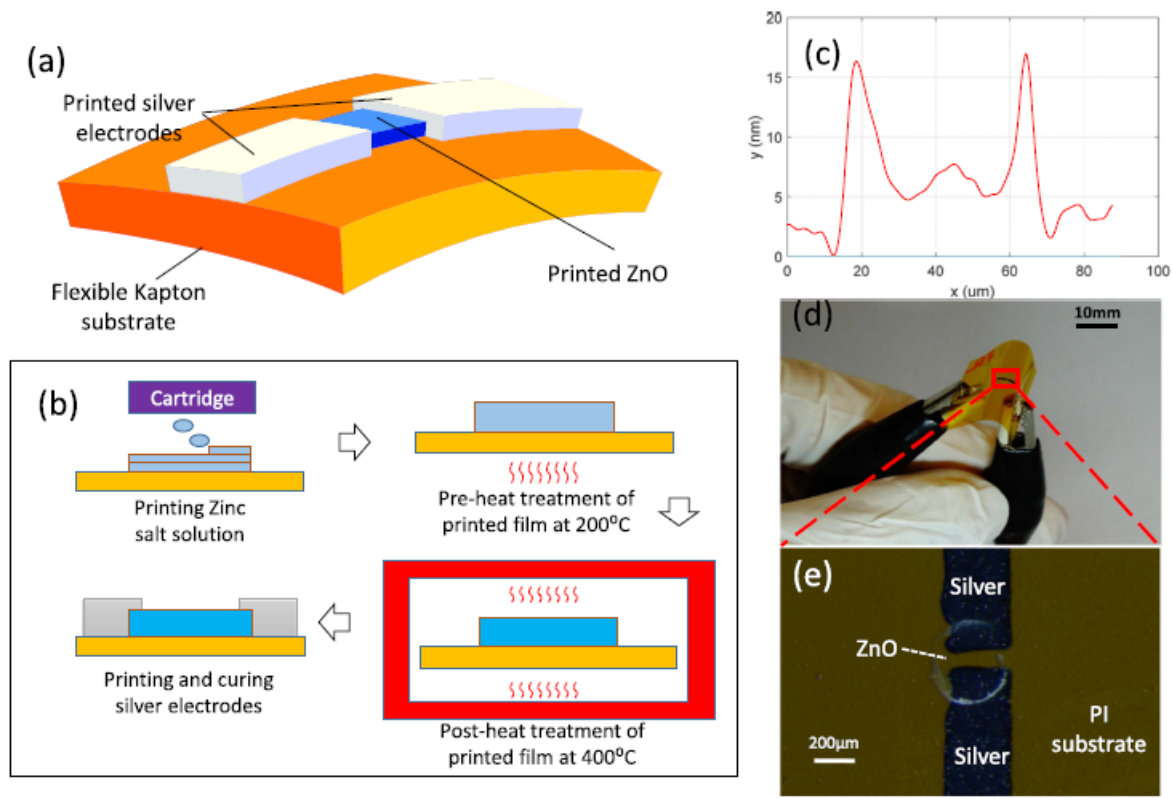

**Figure 14.** (a) Schematic of flexible UV photodetector comprised of inkjet printed ZnO. (b) Fabrication process for the formation of ZnO UV sensor. (c) Height profile of single ZnO droplet. (d) Resulting UV photodetector under mechanical stress by bending. (e) Optical micrograph of ZnO droplet on polyimide substrate.(247)

In work that captures the multi-use nature of ZnO and its applicability to future optoelectronic devices, Tran et al. demonstrate a flexible ZnO UV photodetector formed via inkjet printing of a simple precursor comprised of a simple zinc acetate dihydrate in ethanol.(247) These devices formed at relatively low annealing temperatures of 400 °C display low response times (0.3 s) and high on/off photocurrent ratio. A more architecturally-complex photodetector device was recently reported by Hsu et al. which includes doped ZnO nanowires fabricated via hydrothermal methods.(248) Devices were formed from Y-doped ZnO NWs and subsequently coated in $TiO_2$ deposited using the sol-gel method. Finally, Au nanoparticles were bound to the $TiO_2$ layer via drop casting of chloroauric acid and subsequent annealing in an inert atmosphere. While PL emission demonstrates that $TiO_2$ has the majority impact of the light absorption, fabricated NWs still display high responsivity $1.48 \times 10^{-2}$ A/W. Pickett et al. show hybrid photodetectors and thin film transistors based on the spin coating deposition of zinc acetate dihydrate in 2-methoxyethanol solution.(249) With UV-Ozone cured ZnO thin films, responsivities for hybrid organic-inorganic photodetectors as high as $5.81 \times 10^{-2}$ A/W were achieved.

Recent work exploring all-inkjet printing UV photodetectors have begun exploring the incorporation of nanoparticles (e.g. Ag) into the printed thin films for enhancement of the photodetection capability the semiconductor material(250) (251). Other advances in inkjet printing explore the possibilities of printing heterostructures for improving the performance of ZnO films. Examples include a hybrid graphene/ZnO nanoparticle film, where the graphene acts to improve conductivity and photoresponsivity(252), and p-type NiO/n-type ZnO heterojunction thin films prepared using an all inkjet-based method(253). Additive manufacturing (or 3D printing) is another emerging area for the formation of UV photodetectors, where ZnO particles are typically incorporated into the stock of material (e.g. resin/polymer or filament) used to construct the 3D object to enhance the photo response. Flexible ZnO UV photodetectors were fabricated using filament-based nozzle extrusion 3D printing, consisting of 3D printed layers of polyurethane, poly(vinyl alcohol), Cu-Ag nanowires and ZnO nanoparticles,(254) demonstrating the capability of constructing a photodetector from 3D-printed processes alone.

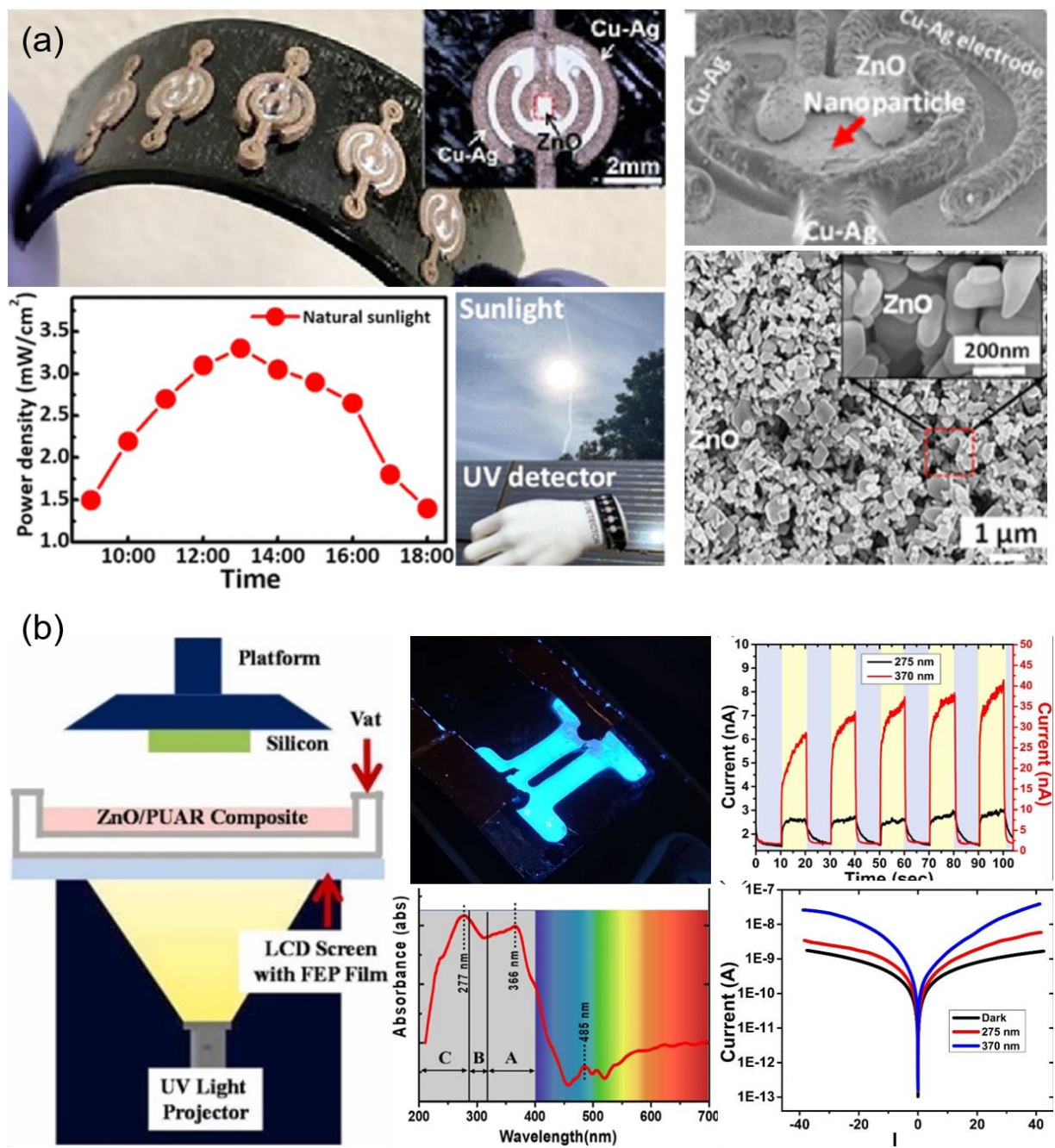

**Figure 15.** (a) 3D printed flexible photodetector with layers deposited through extrusion filament printing utilizing ZnO nanoparticles as UV photosensitive materials(254). (b) A vat polymerization technique used to 3D print a ZnO/polyurethane resin composite with a shown to photo-response to both UV-A (370 nm) and UV-C (275 nm) wavelengths(255).

A wide spectrum photodetector was constructed via a digital light processing 3D printing technique, through which a ZnO/polyurethane acrylate resin was used as the stock material for printing the photodetector(255). The film was tested for photosensitivity in both UV-A (370 nm) and UV-C (275 nm) wavelengths, with an overall broad spectral absorption, and hence photosensitivity, spanning much of the UV region. This process can be seen illustrated in Fig. 15 (b). Similarly, ZnO nanoparticles incorporated into polymeric photoactive materials were used to 3D print a flexible photodetector which could be interfaced to human skin to measure spectral irradiance over a specified time period(256). A 24-hour detection of wavelengths in the region of 310 – 650 nm, corresponding to natural light irradiance, was demonstrated to be possible with these flexible and wearable sensors. These types of new and emerging manufacturing techniques, give a glimpse into the future of photodetector devices and the critical role that ZnO plays in their operation, given its ubiquitous position as a robust and inexpensive semiconductor with a wide bandgap.

**Green Chemistry, ZnO, Clean Energy and Sustainability**

Green chemistry is an ever-expanding area of chemistry with an approach focusing on reducing or removing hazardous materials used in chemical processes. These hazards can be present at various stages of a chemical production process including raw material sourcing, synthesis, processing, application and waste removal. The hazards encompassed by this process include concepts which are harmful to human health, natural environmental resources, ecological habitats and the climate. When speaking of green chemistry, it is common to refer to twelve principle of green chemistry(257) (258) (259), as outlined by Anastas and Werner(260), which can be seen graphically illustrated in Figure 16 (a) (261). As our understanding of the impact of chemical processes increases, it is becoming more and more apparent that some level of intervention is necessary to curb the damage incurred on our environment and health. $CO_2$ emissions, their impact on the global warming and economic incentives to meet lower carbon emissions highlight the need for innovation in $CO_2$ waste management(262) (263) (264).

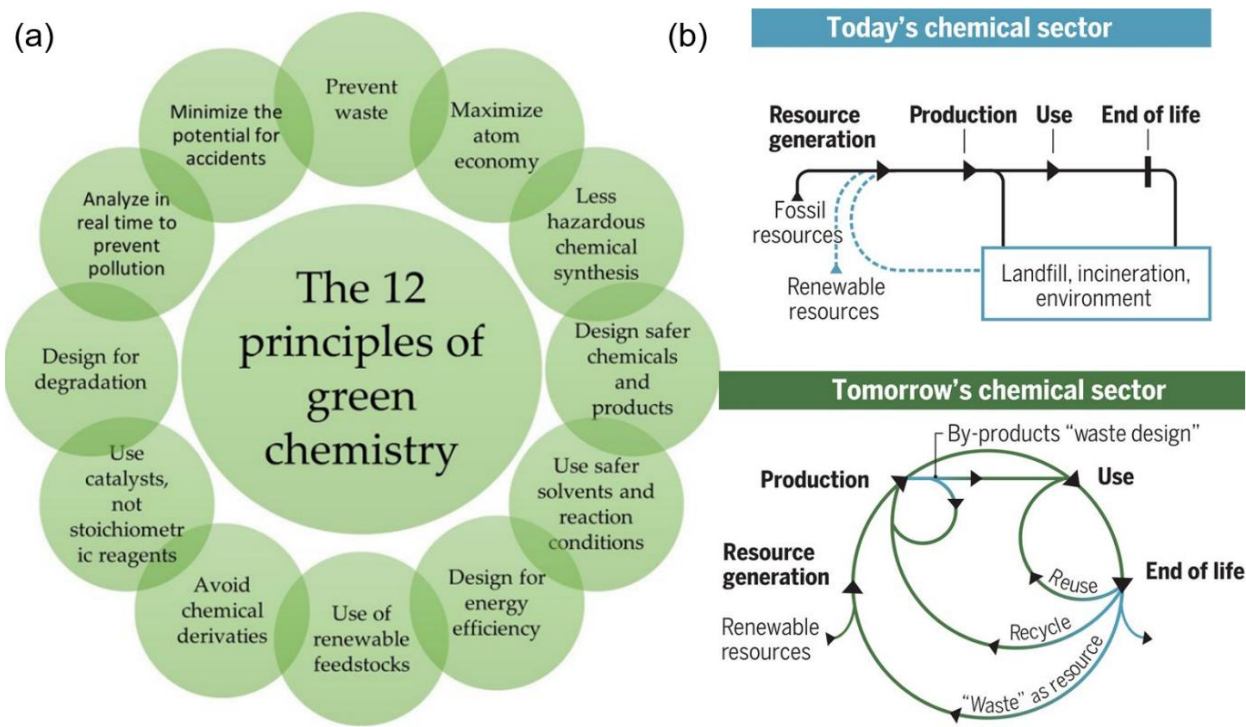

**Figure 16.** (a) A graphically illustration of the 12 principles of green chemistry(261). (b) The cycle life of chemicals processed today versus the ideal cycle created by implementing green chemistry principles(265).

Ubiquitous plastic pollution and the detection of microplastics in the environment outlines how important the effective management of waste is, with many research efforts focusing on metal-free catalysis for plastic breakdown(266) (267) (268). For general chemical processes, toxic metals should be avoided for their impact on the environment and bioaccumulation concerns. Metal elements of Hg, Cd, Cr(VI) and Pb are considered very toxic whereas metals ion of In, Sb, and Bi are considered harmful but less toxic(269) (270). Metals such as Cu, Zn and Fe are considered the least toxic elements. Green chemistry techniques aim to identify switch from toxic metals to less environmentally harmful metals in material selection(271).

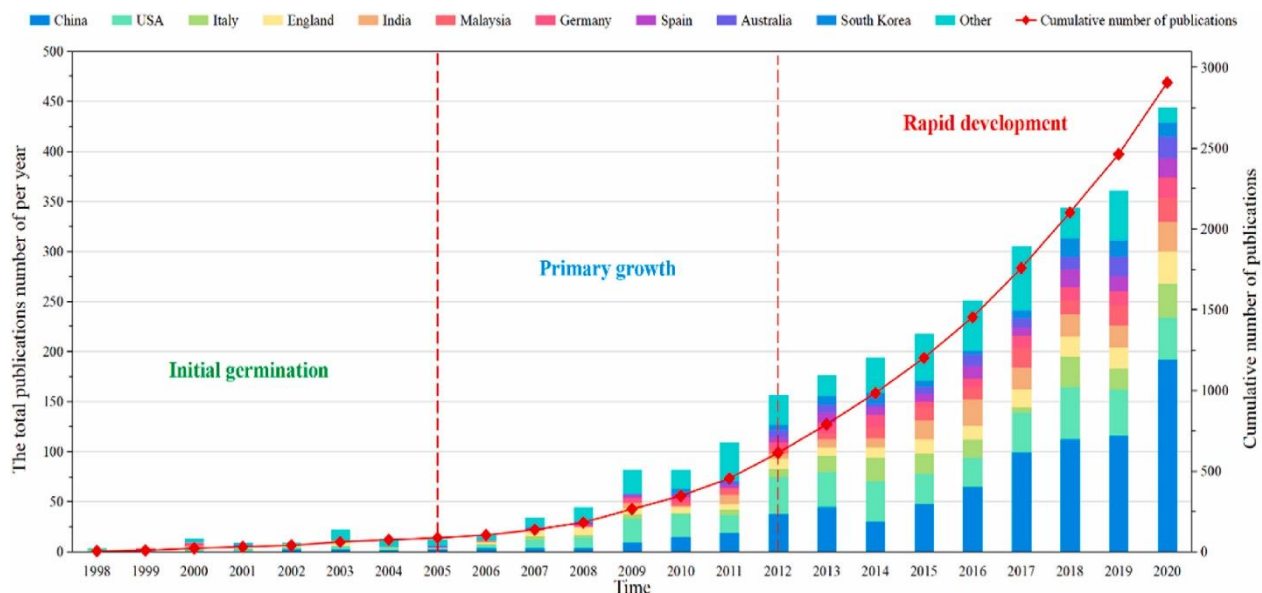

**Figure 17.** Graphic representation showing the number of articles published annually on green energy and environmental technologies, highlighting the growth in the sector over a 22 year period(272).

One of the largest undertakings in the green chemistry discipline is educating and communicating the need for adoption of green chemistry methods in a practical sense(265) (273) (274) (275). The concept of a circular economy is a cornerstone of green chemistry practices; a circular economy aims to balance economic growth, resource sustainability and environmental protection(275) (276). The critical importance of adopting green chemistry methods has even led to calls of just referring to this field as simply "chemistry"(277), given how necessary advancements in this area will be towards damage reduction on environmental resources. This urgency is reflected in the scientific literature with rapid growth in green chemistry publications over the last decade or so, as seen depicted in Figure 17(272). In this work, we place focus on thin film applications with particular attention on ZnO-based materials. Here, we will overview the role of green chemistry in relation to ZnO production and the role of ZnO-based thin film applications in sustainable chemistry.

**Green Chemistry for ZnO Nanoparticle Production**

Green chemistry methods have been used to successfully produce ZnO nanoparticles from a variety of different base material reagents with various particle sizes, morphology and material properties. There are many excellent review articles which report on the abundance of different methods used in the green synthesis of ZnO nanoparticles(278) (279) (280) (281). Presently there are many different biological materials used green ZnO production. Plants (leaves(282), fruits(283), stems(284), roots(285) and flowers(286)), microorganisms (bacteria(287) and fungi(288)) and algae(289) have all been utilized in green synthesis of ZnO nanoparticles. Plant-based synthesis has received a lot of attention in the literature with a plethora of plants explored as reagents for the synthesis reaction. Figure 18 (a) displays various different sources of organic materials used in the green synthesis of ZnO nanoparticles(290).

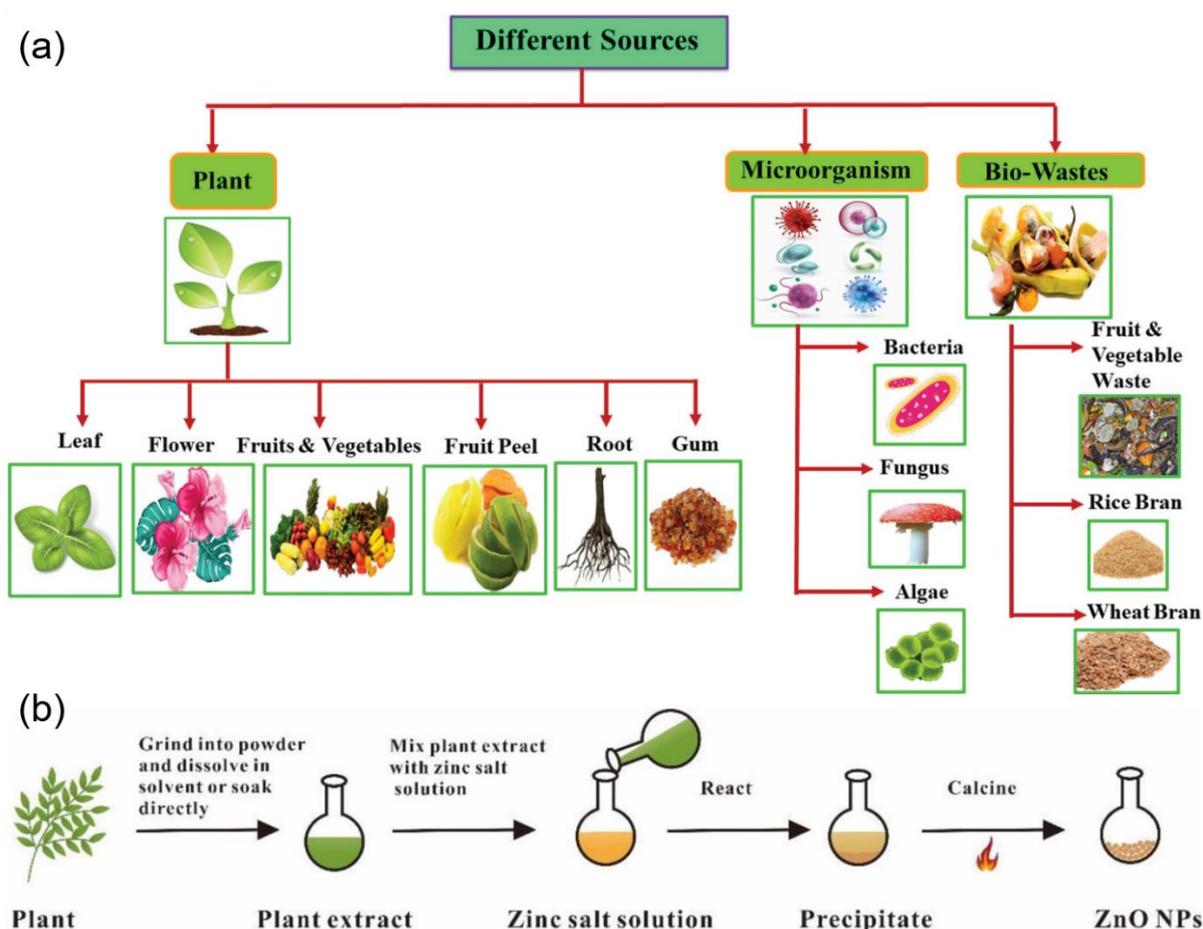

**Figure 18.** (a) Graphic displaying the range of biologically-derived materials used in green synthesis reactions of ZnO nanoparticles(290). (b) A general reaction scheme for the production of ZnO nanoparticles in green chemistry(291).

Reactions involving plants have the advantage of being typically easy to obtain, lower cost, relative abundance for larger scale reactions(292) and usually only require a Zn metal salt solution as a precursor (e.g. zinc acetate, zinc nitrate, zinc chloride, zinc sulfate, etc.)(279) (291). The natural compounds found in plant extracts can act simultaneously as reducing agents for metal ions or metal oxides in precursors and as stabilization or capping agents. The natural compounds involved in the reduction process are typically phytochemicals or antioxidants such as polysaccharides, flavonoids, phenolic compounds, alkaloids, amino acids, terpenoids, tannins and vitamins(278) (291). The process of green synthesis of ZnO nanoparticles follows a general process where the plant material is first thoroughly washed with distilled water, dried, crushed into particulates with a mortar and pestle, dispersal into a suitable solvent, filtration to remove remaining solid organic material, combination with a Zn metal salt precursor and a final calcination of the precipitate formed. This process can be seen depicted in Figure 18 (b)(291).

The properties of the produced ZnO nanoparticles will be dependent on a number of different reaction conditions including choice of plant extract, concentration of extract, metal salt precursor and its concentration, reaction time, pH and calcination time and temperature(281) (290) (291). One study compared the effects of changing the metal-ion precursor used with bay leaves in the green synthesis of ZnO nanoparticles, comparing the effects of zinc acetate versus zinc nitrate precursor(293). The zinc acetate precursor resulted in nanoparticles with bullet-like morphologies with average sizes of 21.49 nm. The zinc nitrate precursor yielded nanoparticles with flower-like morphologies with average sizes of 25.26 nm. In this case, both the morphology and size was influenced by the choice of precursor metal salt. Figure 19 (a) shows the variation in ZnO material produced using different metal salts(293).

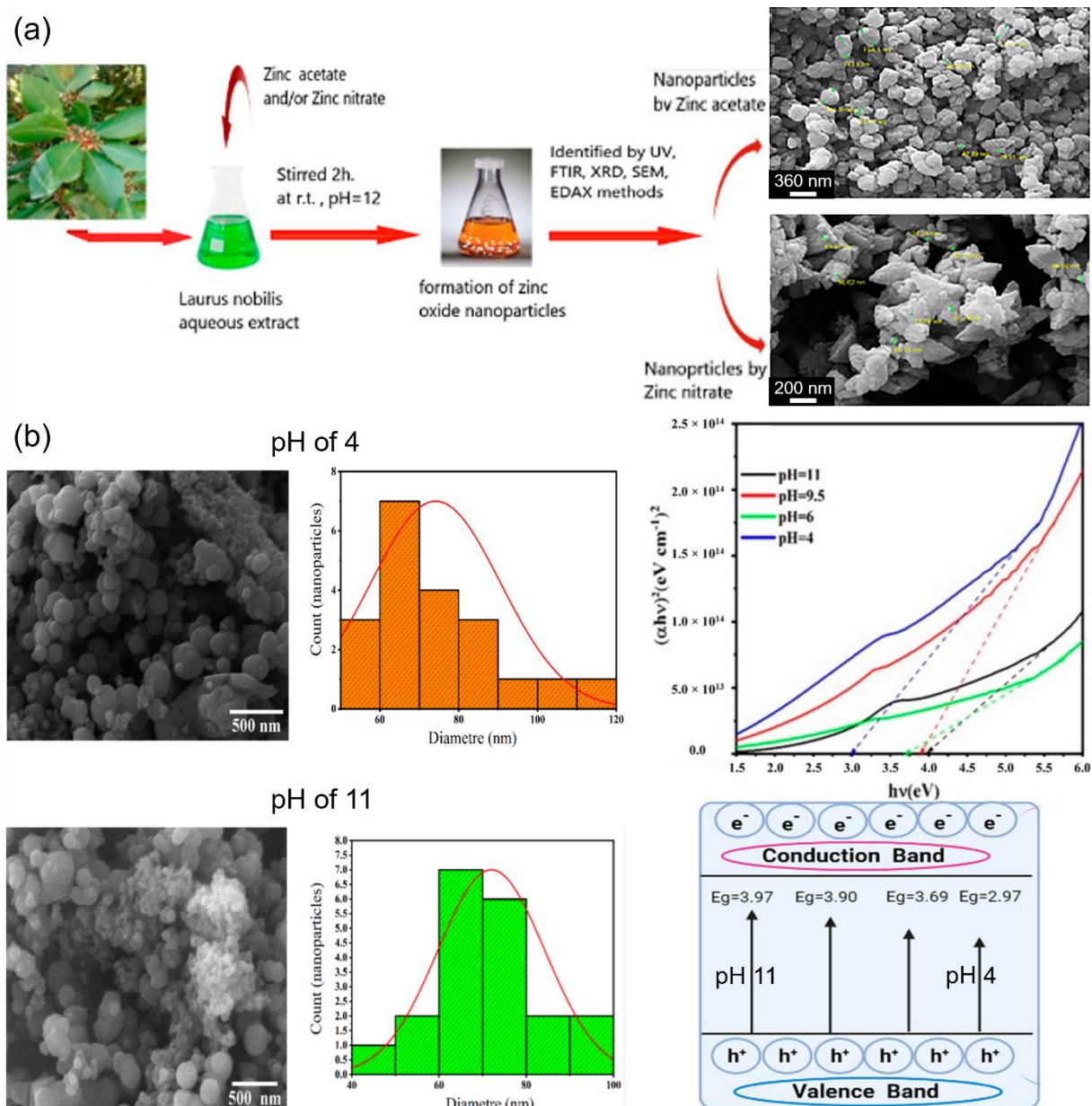

**Figure 19** (a) ZnO nanoparticles synthesized from bay leaves via a green chemistry method exploring the differences in particle morphology using different zinc metal precursors(293). (b) Green chemistry synthesis of ZnO nanoparticles using portulaca oleracea leaves exploring the effect of varying pH values in the reaction on particle morphology, size and bandgap(294).

Another study working with green synthesis of ZnO nanoparticles using portulaca oleracea (pursley) leaves found that the pH of the reaction had an effect on the morphology and associated bandgap of the resultant ZnO nanoparticles produced(294). Basic pH values of 11 resulted in overall smaller average particles of about 22 nm and the widest direct bandgap of about 3.97 eV. Acidic pH values of 4 yielded larger particles with average sizes of about 27 nm with the smallest direct bandgap of 2.97 eV. Figure 19 (b) displays these structures derived by varying pH(294). Studies such as these highlight how the properties of the ZnO material produced can be fine-tuned by varying the reaction conditions used in green synthesis, an important consideration for optoelectronic applications.

**Green/Sustainable ZnO for Optoelectronic Applications**

ZnO nanoparticles manufactured via green chemistry methods have been utilized abundantly in the scientific literature for a number of different applications. There are numerous reports highlighting how green-manufactured ZnO can be used in biological settings with antimicrobial(295) (296), anticancer(297) (298), biosensing(299) (300) and a range of different biomedical(301) (302) (303) applications. Green ZnO has also been extensively explored for wastewater treatment, typically acting as a photocatalyst leveraging the intrinsically wide bandgap of ZnO, for removal of organic dyes(304) (305) (306), organic hydrocarbons in fuel(307), medications such as ibuprofen(306) and paracetamol(308), pesticides(309) and heavy metal ions(310) (311). There is ample precedent in the literature for the versatility and general applicability for green-manufactured ZnO. New implementations are ever-emerging from the diverse range of nanoparticles, and their associated properties, available for the green synthesis of ZnO. Here, we give focus to some of the optoelectronic applications from green ZnO, outlining the role of ZnO in green technology moving forward.

One major field which has shown a strong adoption of green ZnO is dye-sensitized solar cells (DSSCs)(312) (313), with its high electron mobility, conductivity, stability and inexpensive production all contributing to its appeal. ZnO is commonly deployed as the photoanode in DSSCs, typically a semiconducting layer with an associated high surface area onto which the dye molecules are embedded. The electrons generated by the dye molecules under light irradiation are passed into the conduction band of the semiconductor, eventually passing to the counter electrode through the external circuit. Additionally, the ZnO layer in DSSCs is commonly deposited through solution processing techniques such as doctor blading.

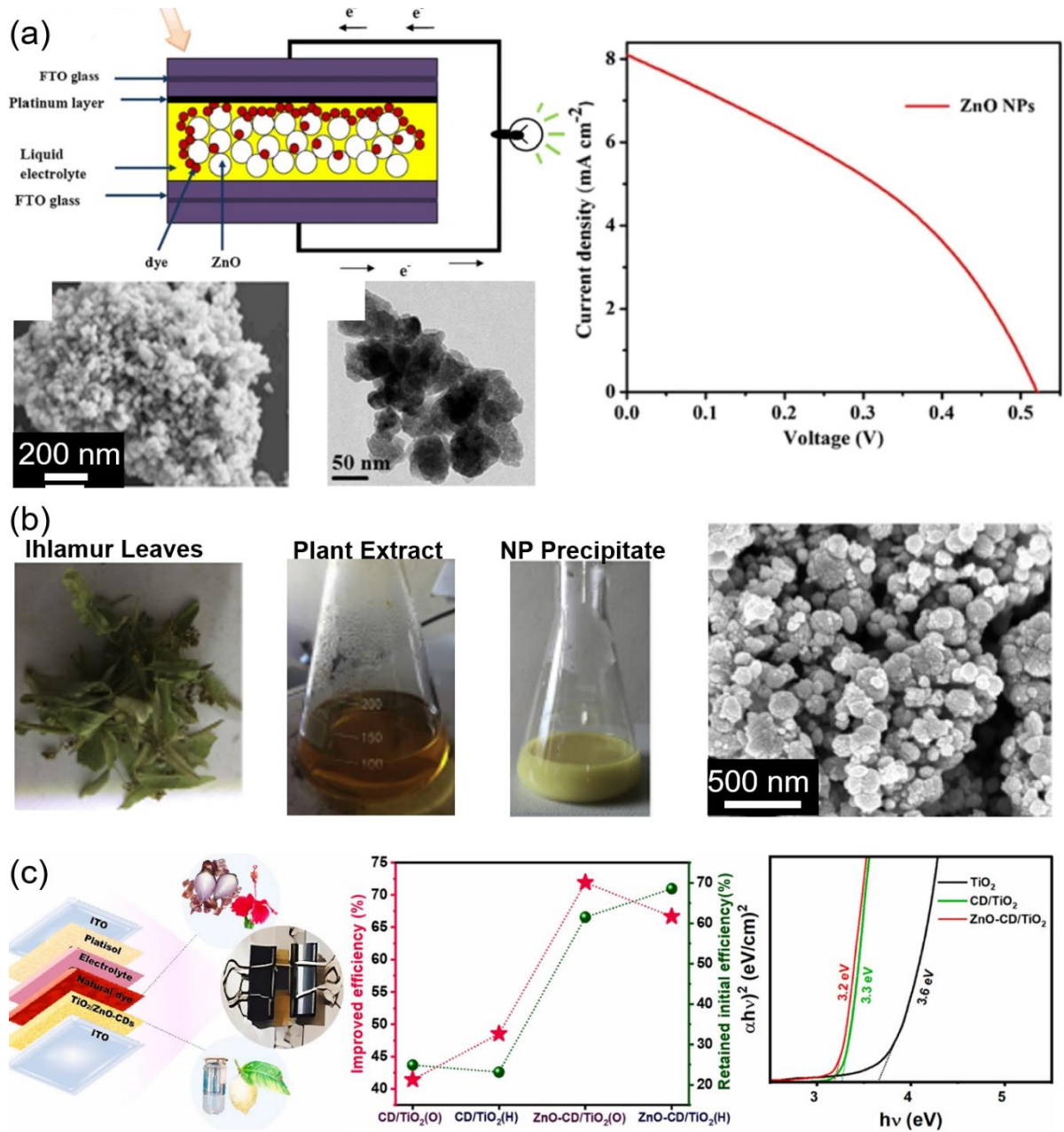

**Figure 20** (a) 50 nm wurtzite ZnO nanoparticles prepared using papaya leaf extract for use in a DSSC(314). (b) Process for preparing 80 nm ZnO nanoparticles using ihamur leaves for green synthesis of a DSSC photoanode(315). (c) Green synthesis using citrus medica fruit extract of a composite hetero structure of carbon dots / ZnO / TiO$_2$ for use with naturally derived sensitizer in a DSSC with the composite structure exhibiting most favourable photo efficiency(316).

Wurtzite ZnO nanoparticles of approximate sizes of 50 nm have been prepared through green synthesis using papaya leaf extract and coated onto a DSSC using a doctor blade method(314). This type of cell reported a conversion efficiency of 1.6 % and a current density of 8.12 mA cm$^{-2}$, as seen depicted in Figure 20 (a). A separate study(315) using the green synthesis of tilia tomentosa leaves for the production of ZnO nanoparticles of approximate sizes 80 nm (see Figure 20 (b)) with a measured bandgap of 3.55 eV deposited into a thin film DSSC layer via a doctor blade method, recorded a conversion efficiency of 1.97 % and a current density of 6.26 mA cm$^{-2}$. Another study examined the effects of different common plant extracts (namely cabbage, onion, carrot and tomato) on the morphology and performance of ZnO nanoparticles used in DSSCs(317). Particle sizes, bandgaps and current densities were all impacted by the plant choice, illustrating the ability of ZnO particle properties to tune the performance of DSSCs.

A common approach in DSSCs is to combine semiconductor materials to achieve better performances through improved electrical conductivity, electron mobility or chemical stability. In general, the heterojunctions formed from composite semiconductor layers also avail of the added benefit of increased charge carrier separation, thus reducing the rate of unwanted charge recombination in the solar cell(318). Hetero-structures of ZnO/TiO$_2$ are commonly investigated for DSSCs in the literature, with researchers attempting to combine the higher chemical stability of TiO$_2$ with the higher electron mobility on ZnO(319) (320) (321). There have been some works which have explored the green synthesis of ZnO/TiO$_2$ hetero structures for use as photocatalysts or as photoanode layers in DSSCs. Hybanthus enneaspermus leaf extract was used to fabricate TiO$_2$ / CuO / ZnO composite nanostructures for use as photocatalysts in rhoadmine dye degradation(322). Recently, carbon dot / ZnO composites were prepared from citrus medica fruit extract and combined with TiO$_2$ nanoparticles to form a ternary material used in DSSCs(316). These structures were used in conjunction with naturally derived dye sensitizers from as Hibiscus rosa-sinensis and Allium Cepa peel for a more sustainable solar cell design. The composite structure outperformed unmodified $TiO_2$ photoanodes in terms of photo conversion efficiency when tested with both natural sensitizers, see Figure 20 (c) for results. These results outline how naturally derived composite heterojunction materials can enhance DSSC performance(316).

Finally, green ZnO has also been utilized as a detector layer in specialised sensor devices with applications hinging on its electronic or optical properties. ZnO nanoparticles fabricated using Camellia japonica leaf extract showed optical sensitivity towards certain heavy metal ions, namely $Ag^+$ and $Li^+$, with clear decreases in absorbance intensity at around 300 nm correlating with increasing metal ion concentration(323). This type of application could be used for selectively testing for specific metal ion contamination in marine water for instance. Figure 21 (a) illustrates the operation of this optical sensor. Another approach leveraged the resistivity response of a composite ZnO / NiO film, prepared via a green synthesis method using Azadirachta Indica leaf extract, to develop a resistance-based sensor for detection of liquefied petroleum gas below the lower exposure limit at room temperature(324). The combination of the n-type ZnO and p-type NiO showed the most effective sensing performance with the largest changes in resistance and highest long-term stability recorded for the composite structure when compared to pure NiO or ZnO. Figure 21 (b) displays the mechanisms by which this resistance sensor operates. Additionally, the sensor film was deposited via spin-coating, a solution processing technique. Recently, a $Ga_2O_3$ / ZnO composite was developed using an Aegle marmelos fruit extract with a recorded bandgap of 3.1 eV and a broad blue photoluminescence emission for an excitation wavelength of 350 nm(325). These specific optical properties of the composite material create an application in latent fingerprint analysis which uses blue light for its ability to interact with chromophores and biomolecules left by organic residues in fingerprints. Figure 21 (c) illustrates this fingerprint detection application across a number of different surfaces.

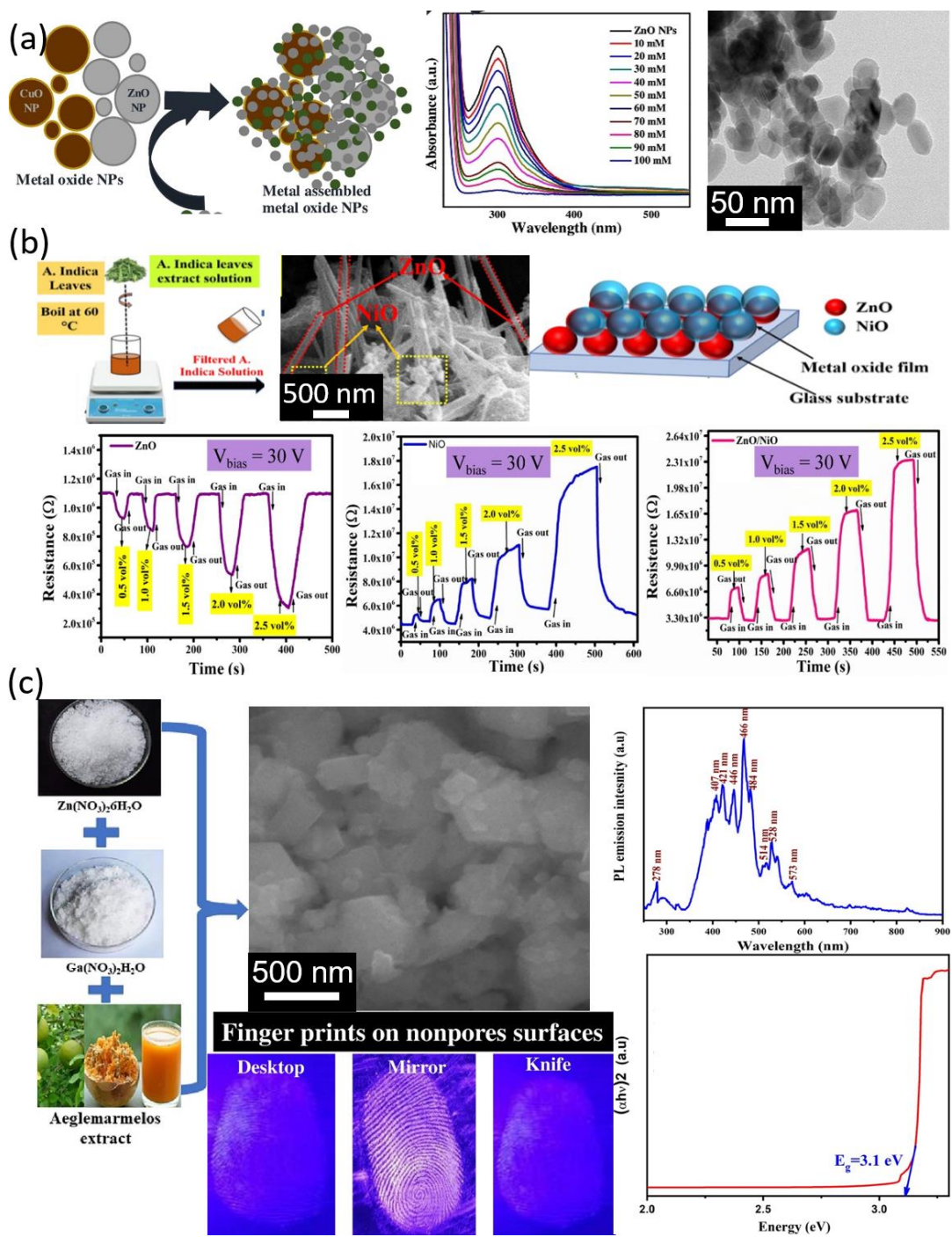

**Figure 21.** (a) Green synthesis ZnO nanoparticles from Camellia japonica leaves for metal ion sensing, showing the change in absorbance for increasing concentration of Ag$^+$ ions in solution(323). (b) ZnO / NiO composite structure prepared using Azadirachta Indica leaf extract, deposited as a thin film for liquefied petroleum gas sensing using the resistance of the film(324). (c) Ga$_2$O$_3$ / ZnO composites prepared from Aegle marmelos fruit extract exhibiting a deep blue photoluminescence response to 350 nm excitation wavelength for applications in latent fingerprint testing, shown for different non porous surfaces(325).

**Sustainable electronics**

More recent efforts involving ZnO and related compounds such as indium zinc oxide (IZO) and others, often in amorphous form, have demonstrated how transistor structures can be made using paper substrates. This type of innovation not only proves how the sustainable materials such as cellulose can in principle be used as gate dielectrics, but that the overall substrate is feasible for paper-type electronic with channel dielectric in thin film form. In the field of memory devices, weight and cost have drastically reduced, but there is demand for lowering these metrics even further. A significant driver for these is the need for recyclability, with sustainability built-in to material choice, application, and how those parameter influence the necessary performance metrics of the technology. Some jurisdictions have requirements for products and component that use plastics, requiring recyclability to some defined degree as a de-facto product requirement. In the microelectronic industry, making memory devices and technologies based on transistor and related concepts, there is an urgent need to find lower cost options that allow fabrication using material that are sustainable, deposited and fabricated at lower temperatures and allow scalability without performance compromise.

For random access memory, non-volatile memory and others, organic transistors and systems have led the way in this regard, but are carrier mobility limited, significantly. If oxide could be used from low temperature deposition onto sustainable and cheaper substrate with natural dielectrics, this would be a significant competitor to organic system, and be more robust in operation. Recently, some groups have tackled the idea of using natural cellulose fibers as a form of paper substate (326). Interestingly, these structure are both the mechanical substrate and the dielectric when paired with n-type ZnO-based oxides. These devices work as write-erase and read field-effect transistors (FETs) that use both IZO and IGZO in some instances (327, 328). Using a representative example, IZO/IGZO FET structures were formed on a resin-fused cellulose fibrous substrate, mimicking paper, which is partially porous and acts as the dielectric. Hydrophobic paper surfaces could be sputter coated with the amorphous oxides with the final device shown in Figure 22. Many other examples take a somewhat similar approach (329-331), but in this case the cycled transfer characteristics show promising read-write on-off states that are relatively stable. The authors at least do address the incipient issue of charge trapping in the fibers, requiring a symmetric erase voltage to that or the 'write' voltage from gate to source to effectively balance of the on-off state and fully 'erase' the memory.

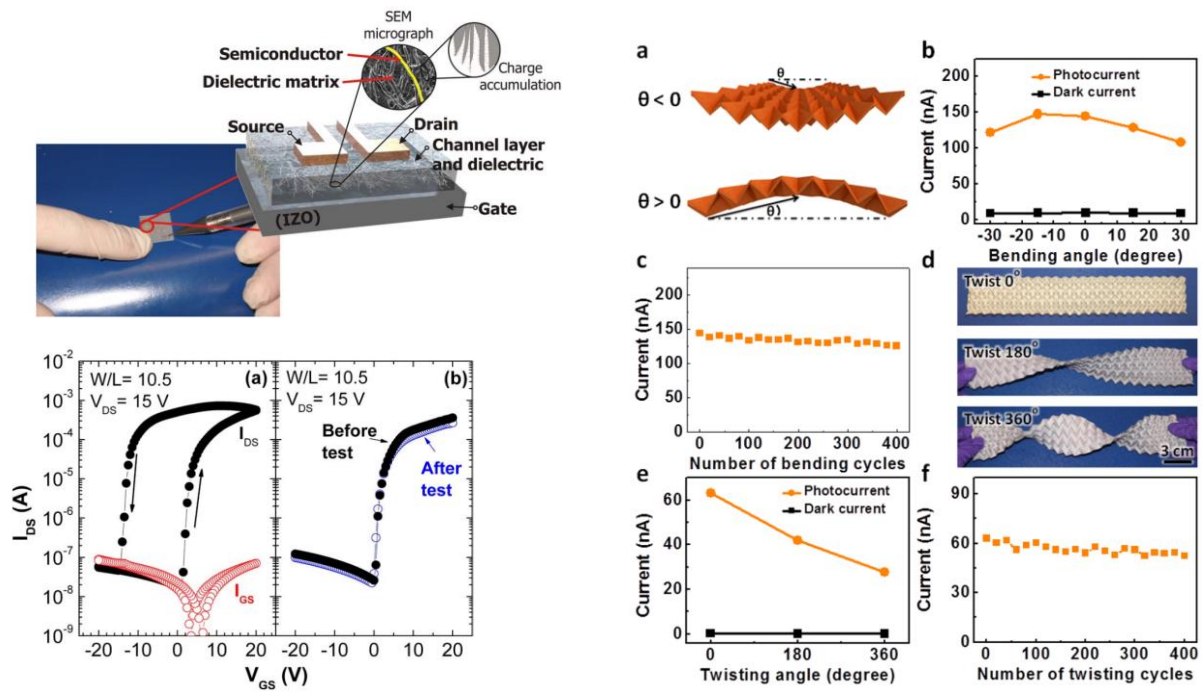

**Figure 22.** (Left) Schematic and photo of the metal oxide on paper device and the fibers that constitute the dielectric structure. Underneath are the transfer characteristics of the write erase read memory-FET showing (a) Double sweep measurements and (b) transfer characteristic before and after the write-erase stress test. (Right) (a) Schematic illustration of origami paper photodetector array (OPPDA) in a concave (bending angle θ < 0°) and convex (bending angle θ > 0°) bending configuration. (b) Bending-angle-dependent photocurrent and dark current of the device under 700% strain. (c) Photocurrent versus bending cycles under UV illumination. (d) Optical photograph of the OPPDA subjected to torsion (angle of twist = 0°, 180°, and 360°). (e) Dark current and photocurrent of the OPPDA versus twist angle measured in the dark and under UV illumination. (f) Photocurrent as a function of the number of twist cycles under UV illumination.

Paper substrates and paper-based optoelectronics are also viable for photodetectors(332-334), where paper-based standard configuration devices have shown promise, as have interesting origami type geometric structures using oxides have been reported. The rationale behind some of these developments has been the internet of things and alternative wearable devices for consumers and for medical or sensor applications. One metric often cited is the 300% stretching capability of certain areas of the skin, which is a consideration for paper, plastic or stretchable polymer-based electronics. In the work of He et al. over the years, along with many others, they developed ZnO-based devices using geometric and folding-capable origami structures that allowed 1000% deformation, 360° twisting rotation and 30° bending angle. While paper-based methods offer significant sustainability benefits compared to plastics and can be modified in-situ during formation, it has been severely limited by a ~5% deformability. But common office paper can be modified and folded using the Miura origami folding methods, a form of geometric accordion-like folding concept that allows significant deformation due to the folded structure. In Figure 22 for example, the rigid folds of the ZnO coated paper maintained tensile stress-free coating as the fold compensated any stretching during twisting. Photodetection remain excellent in this approach since the tessellation of parallelograms allow near omni-directional detectivity, particularly in twisted geometries to give efficient response, even under 700% strain over hundred of twisting cycles.

**Conclusions and outlook**

While there is no shortage of research on ZnO and its related oxides, its useful wide bandgap, n-type conductivity and ease in processing in films and a multitude of nanostructured crystals makes it amenable to unconventional heterogeneous device formulation and investigative devices with alternative substrate types, shapes and composition. The impact of precursor design, process method and formation temperature have a significant impact on the structure type that is formed and their resulting morphological, electrical and optical properties. The ability to modify the optoelectronic properties of ZnO through ion-doping, heterojunction / homojunction composite materials, particle size and film morphology just adds to the appeal of this material for device applications.

ZnO based oxides, including ternary IZO, AZO and quaternary IGZO etc., have proven to be exceptionally useful when paired with other oxides such as $In_2O_3$ as one example. The advent of solution processing of multilayers, heterogenous interfaces and quasi-superlattice structures has provided alternative low temperature solution-processing routes to very high electronic mobility channel materials. This was governed by the quality and band structure difference at the interface between these materials, by methods that are typically less controllable than some sputtering or ALD deposition techniques. The ability of the interface to offer interesting electronic properties makes them potentially interesting optoelectronic materials. Future efforts could consider the leveraging of high mobility interface materials that are very sensitively modified by photon absorption, providing low temperature and lower cost routes to faster switching and responsivities, on substrates that can be folded, stretched, twisted or adapted to confirm to various application form factors. Indium zinc oxide and indium gallium zinc oxide are useful in electronics, where flexibility, transparency, and sustainability are required. Its applications in display technologies, flexible electronics, wearable devices, and even quantum devices are set to grow as more research develops and manufacturers look for sustainable alternatives to ITO. With further improvements in manufacturing processes and device integration, these materials could play a pivotal role in advancing transparent and flexible electronics.

With regards to the future of optoelectronic devices, flexible and wearable technologies are ever-growing industries with projected high market valuations. These types of TFTs require low temperature processing techniques for compatibility with flexible substrates to which solution processed techniques can provide an answer. Considering the future of optoelectronic devices from an environmental perspective, green chemistry methods to synthesize materials are an essential consideration to lower the environmental impact of technological development. Regarding ZnO production specifically, many different green synthesis methods have been explored in the literature with continuous research reports emerging for newer natural reagents and improved ZnO material performance. Certain applications like DSSCs, photocatalysis and light detectors have already adopted the approach of green synthesis of ZnO materials and direct application to an optoelectronic device. Many other areas are likely to follow this trend, of green synthesis to direct application, as sustainable approaches to device fabrication become increasingly necessary.

Material stability, sustainable manufacturing, improving charge carrier mobility and optical transparency are important for these types of oxides, especially where they are used in displays and microLED applications. Control of solution processing will be important in this regard, since roughness, interfacial uniformity, controllable refractive index and optical dispersion need to be controlled carefully. While ZnO has had the widest research investigation, recyclability for ZnO, IZO, IGZO and the many related compounds is an important consideration that needs further work, beyond simply improving the synthetic methods or controlling electrical and optical characteristics at lower temperature. IGZO can be useful in lower power electronics, OLEDs, transparent TFTs and related technologies, having a lower carbon footprint compared to amorphous silicon TFTs for example. But indium recycling will always remain a necessary concern as these materials find new applications.

## Acknowledgements

We gratefully acknowledge the support of the Irish Research Council under a Government of Ireland Postgraduate Scholarship (GOIPG/2014/206), an Advanced Laureate Award (IRCLA/19/118), and a Government of Ireland Postdoctoral Fellowship award (GOIPD/2021/438). Support is acknowledged from Science Foundation Ireland (SFI) through the SFI Technology Innovation and Development Award 2015 under contract 15/TIDA/2893 and by a research grant from SFI under grant Number 14/IA/2581.